\documentclass{aa}  

\bibpunct{(}{)}{;}{a}{}{,}

\usepackage{graphicx, graphics}
\usepackage{txfonts}
\usepackage{longtable}

\begin{document} 

   \title{Revealing the chemical structure of the Class I disc Oph-IRS~67}

   \author{E. Artur de la Villarmois
          \inst{1} \and
          L. E. Kristensen
          \inst{1} \and
          J. K. J{\o}rgensen
          \inst{1}
           }

\institute{Niels Bohr Institute $\&$ Centre for Star and Planet Formation, University of Copenhagen, 
   {\O}ster Voldgade 5--7, 1350 Copenhagen K., Denmark\\
              \email{elizabeth.artur@nbi.ku.dk}
             }

  \abstract
    {Recent results suggest that the first steps towards planet formation may be already taking place in protoplanetary discs during the first 100,000 years after stars form. It is therefore crucial to unravel the physical and chemical structures of such discs in their earliest stages while they are still embedded in their natal envelopes and compare them with more evolved systems.}
    {The purpose of this paper is to explore the structure of a line-rich Class I protobinary source, Oph-IRS 67, and analyse the differences and similarities with Class 0 and Class II sources.}
    {We present a systematic molecular line study of IRS 67 with the Submillimeter Array (SMA) on 1--2$\arcsec$ (150--300~AU) scales. The wide instantaneous band-width of the SMA observations ($\sim$30~GHz) provide detections of a range of molecular transitions that trace different physics, such as CO isotopologues, sulphur-bearing species, deuterated species, and carbon-chain molecules.}
    {We see significant differences between different groups of species. For example, the CO isotopologues and sulphur-bearing species show a rotational profile and are tracing the larger-scale circumbinary disc structure, while CN, DCN, and carbon-chain molecules peak at the southern edge of the disc at blue-shifted velocities. In addition, the cold gas tracer DCO$^{+}$ is seen beyond the extent of the circumbinary disc.}    
    {The detected molecular transitions can be grouped into three main components: cold regions far from the system, the circumbinary disc, and a UV-irradiated region likely associated with the surface layers of the disc that are reached by the UV radiation from the sources. The different components are consistent with the temperature structure derived from the ratio of two H$_{2}$CO transitions, that is, warm temperatures are seen towards the outflow direction, lukewarm temperatures are associated with the UV-radiated region, and cold temperatures are related with the circumbinary disc structure. The chemistry towards IRS 67 shares similarities with both Class 0 and Class II sources, possibly due to the high gas column density and the strong UV radiation arising from the binary system. IRS 67 is, therefore, highlighting the intermediate chemistry between deeply embedded sources and T-Tauri discs.}


   \maketitle

\section{Introduction}

Protoplanetary discs are a crucial stage between warm cores and the formation of planetesimals and planets, and the final composition of planets likely depends on the chemical processing within the disc. The chemical complexity of the disc is established by the material that is incorporated from the inner envelope and by physical processes that promotes a chemical reset \citep[e.g. outflows, accretion shocks, UV radiation field;][]{Bergin2003, Ceccarelli2004, Herbst2009, Sakai2013, Miura2017}. The detail of how and when the disc forms are still not well understood. In particular, the physics and chemistry of the innermost regions are challenging to observe and study, as they are embedded within a large amount of gas and dust. It is, therefore, essential to study and understand the first stages of discs formation. This paper presents an extensive Submillimeter Array (SMA) survey of the molecular line emission on a few hundred AU scales towards the Class I protobinary system Oph-IRS 67, and investigates the physical and chemical signatures associated with the system.   

The complex structure of embedded discs is associated with a huge range of temperatures ($\sim$10$-$1000~K) and densities \citep[10$^{5}$$-$10$^{13}$~cm$^{-3}$;][]{Herbst2009, vanDishoeck2018}, making molecules excellent diagnostics of the physical conditions and processes. Chemical surveys towards deeply embedded sources show the potential of the chemistry to shed light on the physical structure of these regions \citep[e.g.][]{Jorgensen2016, Lefloch2018, Yoshida2019}. Also, the chemical inventories and initial conditions of young discs can then be compared with discs around more evolved (Class II or T Tauri) young stars \citep[e.g.][]{Oberg2011b, Kastner2018}. 

Class I discs represent the bridge between deeply embedded Class 0 sources and the emergence of planet-forming discs, that is, Class II sources. Recent high-sensitivity and spatial resolution observations suggest that planet formation starts at early stages (Class I), based on the evidence of considerable grain growth \citep[e.g.][]{ALMA2015, Harsono2018}. The study of Class I discs is, therefore, essential in order to constrain the first steps of planet formation and to analyse the chemical evolution of discs at different stages. The observations of Class I discs are challenging since they are small in size \citep[$\sim$100~AU;][]{Harsono2014, Yen2015} and other components, such as inner envelope and outflow material, can contribute to the emission on small scales. 

The Class I protobinary system IRS 67 is located in the Ophiuchus star forming region and belongs to the L1689 cloud. It has been associated with a large-scale outflow structure \citep[$\gtrsim$1000~AU;][]{Bontemps1996}, bolometric temperature (\textit{T$_\mathrm{bol}$}) of 130~K, and bolometric luminosity (\textit{L$_\mathrm{bol}$}) of 4.0~L$_{\odot}$ \citep{Dunham2015}. The binarity of the system was proven by \cite{McClure2010} at infrared wavelengths, where a separation of 0$\farcs$6 \citep[$\sim$90~AU for a distance of 151.2~pc;][]{OrtizLeon2017} was found between the sources. More recently, a Keplerian circumbinary disc has been associated to IRS 67 with an extent of $\sim$620~AU \citep{Artur2018}. In comparison with other Class I sources, IRS 67 shows a particular rich chemistry and bright emission of c-C$_{3}$H$_{2}$ \citep{Artur2018, Artur2019}, a molecule commonly associated with photon-dominated regions \citep[PDRs;][]{Guzman2015, Murillo2018}. 

This paper presents an SMA chemical survey towards the Class I proto-binary system IRS 67 and is laid out as follows. Section 2 presents the observational details, data calibration, and spectral coverage. The continuum emission, detected molecular transitions, contour maps, moment 0 maps, and spectral features are described in Sect. 3. We discuss the chemical and physical structure of IRS 67 in Sect. 4, together with an analysis of the non-detected molecular lines, and a comparison with other stages of low-mass star formation. Finally, the main results are summarised in Sect. 5.

\section{Observations}

We observed IRS 67 using the Submillimeter Array (SMA), in the extended and compact configurations on 2017 February 24 and 2017 April 21, respectively (program code: 2016B-S0004; PI: Elizabeth Artur de la Villarmois). The antenna configuration provided projected baselines between 10 and 188~metres. A pointing centre of $\alpha$ = 16$^\mathrm{h}$32$^\mathrm{m}$00$\fs$98, $\delta$ = $-$24$\degr$56$\arcmin$43$\farcs$4 (J2000) was used, corresponding to the peak of the continuum emission at 0.87~mm \citep{Artur2018}. The observations covered a frequency range of $\sim$30~GHz, between 214.3 and 245.6~GHz (Table~\ref{table:observations}) with a spectral resolution of 139.64~kHz (0.18~km~s$^{-1}$).

The calibration and imaging were done in CASA\footnote{\tt http://casa.nrao.edu/} \citep{McMullin2007}. For the extended track, the complex gains were calibrated through observations of the quasar 1626-298, passband calibration on 3c273, and flux calibration on Ganymede. For the compact track, the gain, passband, and flux calibrations were implemented by observing the quasar 1626-298, 3c454.3, and Callisto, respectively. The extended and compact observations were combined and a Briggs weighting with robustness parameter of 1 was applied to the visibilities. The resulting dataset has a typical beam size of 1$\farcs$7~$\times$1$\farcs$4 ($\sim$260~$\times$~210 AU) and a maximum angular scale of $\sim$25$\arcsec$. The continuum and line \textit{rms} are given in Table~\ref{table:observations}.

\begin{table*}[t]
        \caption{Summary of the observations.}
        \label{table:observations}
        \centering
        \begin{tabular}{l l c c c}
                \hline\hline
                Receiver        	& Sideband      	& Frequency range               	& Continuum \textit{rms}  		& Spectral \textit{rms}                 		\\
                                		&                       &  (GHz)                                	& (Jy beam$^{-1}$)                	& (Jy beam$^{-1}$ channel) $^{(a)}$       	\\
                \hline
                230             	& lower         	& 214.3 -       222.3           		& 0.001                           		& 0.08                                         		\\
                240             	& lower         	& 221.6 -       229.6           		& 0.003                           		& 0.15                                         		\\
                230             	& upper        	& 230.3 -       238.3           		& 0.001                          		& 0.10                                         		\\
                240            	 & upper         	& 237.6 -       245.6           		& 0.004                           		& 0.17                                          		\\
                \hline
        \end{tabular}
        \tablefoot{$^{(a)}$ The channel width is 0.18~km~s$^{-1}$.}
\end{table*}

\section{Results}

\subsection{Continuum}

\begin{table*}[t]
        \caption{Results of 2D Gaussian fits towards the continuum peaks.}
        \label{table:fluxes}
        \centering
        \begin{tabular}{l l c c c c}
                \hline\hline
                Receiver        	& Sideband      	& Size $^{(a)}$                                                   	& PA                                      	& Integrated flux               	& Intensity peak                 	\\
                                		&                       & ($\arcsec$)                                                   	& ($\degr$)                       		& (Jy)                          	& (Jy beam$^{-1}$)                    \\
                \hline
                230             	& lower         	& 1.4~$\pm$~0.1~$\times$~0.7~$\pm$~0.2  	& 38~$\pm$~9                      	& 0.087~$\pm$~0.004     	& 0.056~$\pm$~0.002            	\\
                240             	& lower         	& 1.4~$\pm$~0.2~$\times$~1.0~$\pm$~0.4  	& \quad3~$\pm$~44         		& 0.097~$\pm$~0.007     	& 0.064~$\pm$~0.003             	\\
                230             	& upper         	& 1.2~$\pm$~0.1~$\times$~0.6~$\pm$~0.1  	& 41~$\pm$~6                      	& 0.105~$\pm$~0.003     	& 0.071~$\pm$~0.002            	\\
                240             	& upper         	& 1.6~$\pm$~0.3~$\times$~1.3~$\pm$~0.5  	& \enspace24~$\pm$~81     	& 0.130~$\pm$~0.012     	& 0.068~$\pm$~0.005            	\\
                \hline
        \end{tabular}
        \tablefoot{$^{(a)}$ Deconvolved size (FWHM).}
\end{table*}

The continuum emission for each receiver and sideband is fitted with two-dimensional (2D) Gaussians in the image plane, and the resulting deconvolved size, position angle, integrated flux, and intensity peak are listed in Table~\ref{table:fluxes}. The continuum emission at 0.87~mm, previously detected with ALMA \citep{Artur2018}, showed the contribution of both protostellar sources and a circumbinary disc, associated with a position angle of 54$\degr$, an extent of $\sim$620 AU, and a total integrated flux (both sources and disc) of 0.30~Jy. 

In the Rayleigh-Jeans limit of the spectrum, the sub-millimeter flux (\textit{F$_\mathrm{\nu}$}) has a power-law dependence with the frequency, i.e. \textit{F$_\mathrm{\nu}$}~$\propto$~$\nu$$^{\alpha}$, where $\alpha$ is the slope of the sub-mm spectral energy distribution (SED) and is related with the dust opacity spectral index ($\beta$) by $\alpha$~=~$\beta$~+~2 \citep[e.g.][]{Beckwith1991, Testi2014}. By fitting a power law distribution to the integrated fluxes from Table~\ref{table:fluxes} and the 0.87~mm value from \cite{Artur2018}, an $\alpha$ index of 2.8~$\pm$~0.3 is found, and $\beta$~=~0.8~$\pm$~0.3. The typical $\beta$-value of the interstellar medium (ISM) is $\sim$1.7 \citep{Natta2007}, therefore, a lower value suggests grain growth or optically thick dust emission. This is consistent with the results from \cite{Jorgensen2007}, who find an $\alpha$ index of between two and three for a sample of Class 0 protostars.

\begin{table*}[t]
        \caption{Parameters of the detected molecular transitions.}
        \label{table:detected_molecules}
        \centering
        \begin{tabular}{l l c c r l}
                \hline\hline
                Species                      & Transition                            						& Frequency $^{(a)}$	& \textit{A$_{ij}$} $^{(a)}$	& \textit{E$_\mathrm{u}$} $^{(a)}$  	& \textit{n$_\mathrm{crit}$} $^{(b)}$    \\
                                                	&                                               						& (GHz)                   		& (s$^{-1}$)                    	&  (K)                                          	& (cm$^{-3}$)                                     \\
                \hline
                \multicolumn{6}{c}{CO- species}                                                                                                                                                                                                                 									\\
                \hline
                CO                             	& \textit{J}=2$-$1								& 230.53800       	& 6.92 $\times$ 10$^{-7}$      	& 17            					& 2.2 $\times$ 10$^{4}$            		\\
                $^{13}$CO               	& \textit{J}=2$-$1                                                               	& 220.39868       	& 6.07 $\times$ 10$^{-7}$      	& 16            					& 2.0 $\times$ 10$^{4}$             	\\
                C$^{18}$O               	& \textit{J}=2$-$1                                                                	& 219.56035       	& 6.01 $\times$ 10$^{-7}$      	& 16            					& 1.9 $\times$ 10$^{4}$             	\\
                C$^{17}$O               	& \textit{J}=2$-$1                                                               	& 224.71439       	& 6.42 $\times$ 10$^{-7}$       	& 16            					& 2.1 $\times$ 10$^{4}$             	\\
                H$_{2}$CO               	& 3$_{0,3}$$-$2$_{0,2}$ (para)                                       	& 218.22219       	& 2.82 $\times$ 10$^{-4}$       	& 21            					& 3.4 $\times$ 10$^{6}$              	\\
                H$_{2}$CO               	& 3$_{2,2}$$-$2$_{2,1}$ (para)                                          	& 218.47563       	& 1.58 $\times$ 10$^{-4}$       	& 68            					& 1.4 $\times$ 10$^{6}$              	\\
                H$_{2}$CO               	& 3$_{2,1}$$-$2$_{2,0}$ (para)                                          	& 218.76007       	& 1.58 $\times$ 10$^{-4}$       	& 68            					& 3.2 $\times$ 10$^{6}$                	\\
                H$_{2}$CO               	& 3$_{1,2}$$-$2$_{1,1}$ (ortho)                                         	& 225.69778       	& 2.75 $\times$ 10$^{-4}$       	& 33            					& 5.7 $\times$ 10$^{6}$                	\\
                \hline
                \multicolumn{6}{c}{Deuterated species}                                                                                                                                                                                                         								 \\
                \hline
                DCO$^{+}$               	& \textit{J}=3$-$2                                                                	& 216.11258       	& 2.38 $\times$ 10$^{-3}$       	& 21            					& 7.0 $\times$ 10$^{6}$               	\\
                DCN                         	& 3$-$2                                                                                & 217.23854      	& 4.57 $\times$ 10$^{-4}$        	& 21            					& 2.5 $\times$ 10$^{7}$           		\\
                DNC                        	& 3$-$2                                                                                & 228.91049      	& 5.62$\times$ 10$^{-4}$         	& 22            					& 4.9 $\times$ 10$^{6}$           		\\
                \hline
                \multicolumn{6}{c}{S- species}                                                                                                                                                                                       					                                  			\\
                \hline
                CS                            	& 5$-$4                                                                                & 244.93556       	& 2.95 $\times$ 10$^{-4}$       	& 35            					& 8.7 $\times$ 10$^{6}$           		\\
                C$^{34}$S               	& 5$-$4                                                                                & 241.01609       	& 2.75 $\times$ 10$^{-4}$         	& 28           					& 8.0 $\times$ 10$^{6}$           		\\
                $^{13}$CS               	& 5$-$4                                                                               	& 231.22069       	& 2.51 $\times$ 10$^{-4}$         	& 33           					& 7.4 $\times$ 10$^{6}$          		\\
                SO                            	& 5$_{5}$$-$4$_{4}$                                                          	& 215.22065       	& 1.20 $\times$ 10$^{-4}$         	& 44           					& 1.8 $\times$ 10$^{8}$ $^{(c)}$    	\\
                SO                            	& 6$_{5}$$-$5$_{4}$                                                        	& 219.94944       	& 1.35 $\times$ 10$^{-4}$         	& 35           				 	& 2.5 $\times$ 10$^{8}$ $^{(c)}$    	\\
                \hline
                \multicolumn{6}{c}{N- species}                                                                                                                                                                                                                         								\\
                \hline
                CN $^{(d)}$                & \textit{N}=2$-$1, \textit{J}=3/2$-$3/2, \textit{F}=1/2$-$1/2    & 226.28742     	& 1.02 $\times$ 10$^{-5}$        	& 16              					& 6.1 $\times$ 10$^{4}$         		\\
                CN                            	& \textit{N}=2$-$1, \textit{J}=3/2$-$3/2, \textit{F}=5/2$-$5/2    & 226.35987     	& 1.62 $\times$ 10$^{-5}$         	& 16             					& 5.8 $\times$ 10$^{5}$         		\\
                CN                            	& \textit{N}=2$-$1, \textit{J}=3/2$-$1/2, \textit{F}=1/2$-$3/2    & 226.61657     	& 1.07 $\times$ 10$^{-5}$		& 16              					& 1.8 $\times$ 10$^{5}$         		\\
                CN                            	& \textit{N}=2$-$1, \textit{J}=3/2$-$1/2, \textit{F}=3/2$-$3/2    & 226.63219     	& 4.27 $\times$ 10$^{-5}$		& 16              					& 2.6 $\times$ 10$^{5}$         		\\
                CN                             & \textit{N}=2$-$1, \textit{J}=3/2$-$1/2, \textit{F}=5/2$-$3/2    & 226.65956     	& 9.55 $\times$ 10$^{-5}$		& 16              					& 7.0 $\times$ 10$^{5}$        		\\
                CN                             & \textit{N}=2$-$1, \textit{J}=3/2$-$1/2, \textit{F}=1/2$-$1/2    & 226.66369     	& 8.51 $\times$ 10$^{-5}$		& 16              					& 6.0 $\times$ 10$^{5}$         		\\
                CN                             & \textit{N}=2$-$1, \textit{J}=3/2$-$1/2, \textit{F}=3/2$-$1/2    & 226.67931    	& 5.25 $\times$ 10$^{-5}$       	& 16              					& 3.8 $\times$ 10$^{5}$         		\\
                CN                             & \textit{N}=2$-$1, \textit{J}=5/2$-$3/2, \textit{F}=5/2$-$3/2    & 226.87419     	& 9.55 $\times$ 10$^{-5}$     	& 16              					& 2.4 $\times$ 10$^{6}$         		\\
                CN                             & \textit{N}=2$-$1, \textit{J}=5/2$-$3/2, \textit{F}=7/2$-$5/2    & 226.87478     	& 1.15 $\times$ 10$^{-4}$  	& 16              					& 5.1 $\times$ 10$^{7}$         		\\
                CN                             & \textit{N}=2$-$1, \textit{J}=5/2$-$3/2, \textit{F}=3/2$-$1/2    & 226.87590     	& 8.51 $\times$ 10$^{-5}$      	& 16            					& 1.2 $\times$ 10$^{6}$        		\\
                CN                             & \textit{N}=2$-$1, \textit{J}=5/2$-$3/2, \textit{F}=3/2$-$3/2    & 226.88742     	& 2.75 $\times$ 10$^{-5}$     	& 16            					& 9.5 $\times$ 10$^{5}$         		\\
                CN                             & \textit{N}=2$-$1, \textit{J}=5/2$-$3/2, \textit{F}=5/2$-$5/2    & 226.89213     	& 1.82 $\times$ 10$^{-5}$         	& 16              					& 5.4 $\times$ 10$^{5}$         		\\
                $^{13}$CN               	& \textit{N}=2$-$1, \textit{J}=5/2$-$3/2, \textit{F}=4$-$3          & 217.46715     	& 1.01 $\times$ 10$^{-4}$        	& 16              					&                                               		\\
                \hline
                \multicolumn{6}{c}{Carbon-chain species}                                                                                                                                                                                                               							\\
                \hline
                c-C$_{3}$H$_{2}$   	& 3$_{3,0}$$-$2$_{2,1}$ (ortho)					& 216.27876       	& 2.57 $\times$ 10$^{-4}$   	& 19            					& 2.0 $\times$ 10$^{8}$           		\\
                c-C$_{3}$H$_{2}$    	& 6$_{0,6}$$-$5$_{1,5}$ (para)                                          	& 217.82215       	& 5.37 $\times$ 10$^{-4}$   	& 39            					& 8.7 $\times$ 10$^{7}$           		\\
                c-C$_{3}$H$_{2}$    	& 5$_{1,4}$$-$4$_{2,3}$ (ortho)                                         	& 217.94005      	& 3.98 $\times$ 10$^{-4}$  	& 35            					& 6.3 $\times$ 10$^{7}$           		\\
                c-C$_{3}$H$_{2}$    	& 5$_{2,4}$$-$4$_{1,3}$ (para)                         			& 218.16046      	& 4.07 $\times$ 10$^{-4}$        	& 35            					& 9.0 $\times$ 10$^{7}$           		\\
                c-C$_{3}$H$_{2}$    	& 7$_{2,6}$$-$7$_{1,7}$ (para)                                		& 218.73273      	& 8.91 $\times$ 10$^{-5}$      	& 61            					& 2.9 $\times$ 10$^{7}$           		\\
                c-C$_{3}$H$_{2}$    	& 4$_{3,2}$$-$3$_{2,1}$ (ortho)                                         	& 227.16914      	& 3.09 $\times$ 10$^{-4}$      	& 29            					& 2.8 $\times$ 10$^{7}$           		\\
                HC$_{3}$N               	& \textit{J}=24$-$23                                                           	& 218.32472      	& 8.32 $\times$ 10$^{-4}$     	& 131   						& 4.7 $\times$ 10$^{7}$                	\\
                HC$_{3}$N               	& \textit{J}=25$-$24                                                           	& 227.41891      	& 9.33 $\times$ 10$^{-4}$       	& 142   						& 1.2 $\times$ 10$^{7}$                	\\
                HC$_{3}$N               	& \textit{J}=26$-$25                                                          	& 236.51279      	& 1.05 $\times$ 10$^{-3}$         	& 153   						& 3.2 $\times$ 10$^{7}$                	\\      
                \hline
        \end{tabular}
        \tablefoot{$^{(a)}$  Values from the CDMS database \citep{Muller2001}. $^{(b)}$ Calculated values for a collisional temperature of 30~K and collisional rates from the Leiden Atomic and Molecular Database \citep[LAMDA;][]{Schoier2005}. $^{(c)}$ For a collisional temperature of 60~K. $^{(d)}$ CN transitions with \textit{A$_{ij}$} higher than 1~$\times$~10$^{-5}$~s$^{-1}$. The references for collisional rates of specific species are presented in Table~\ref{table:non_detected_molecules}.}
\end{table*}

\begin{figure*}[h]
   \centering
      \includegraphics[width=.98\textwidth]{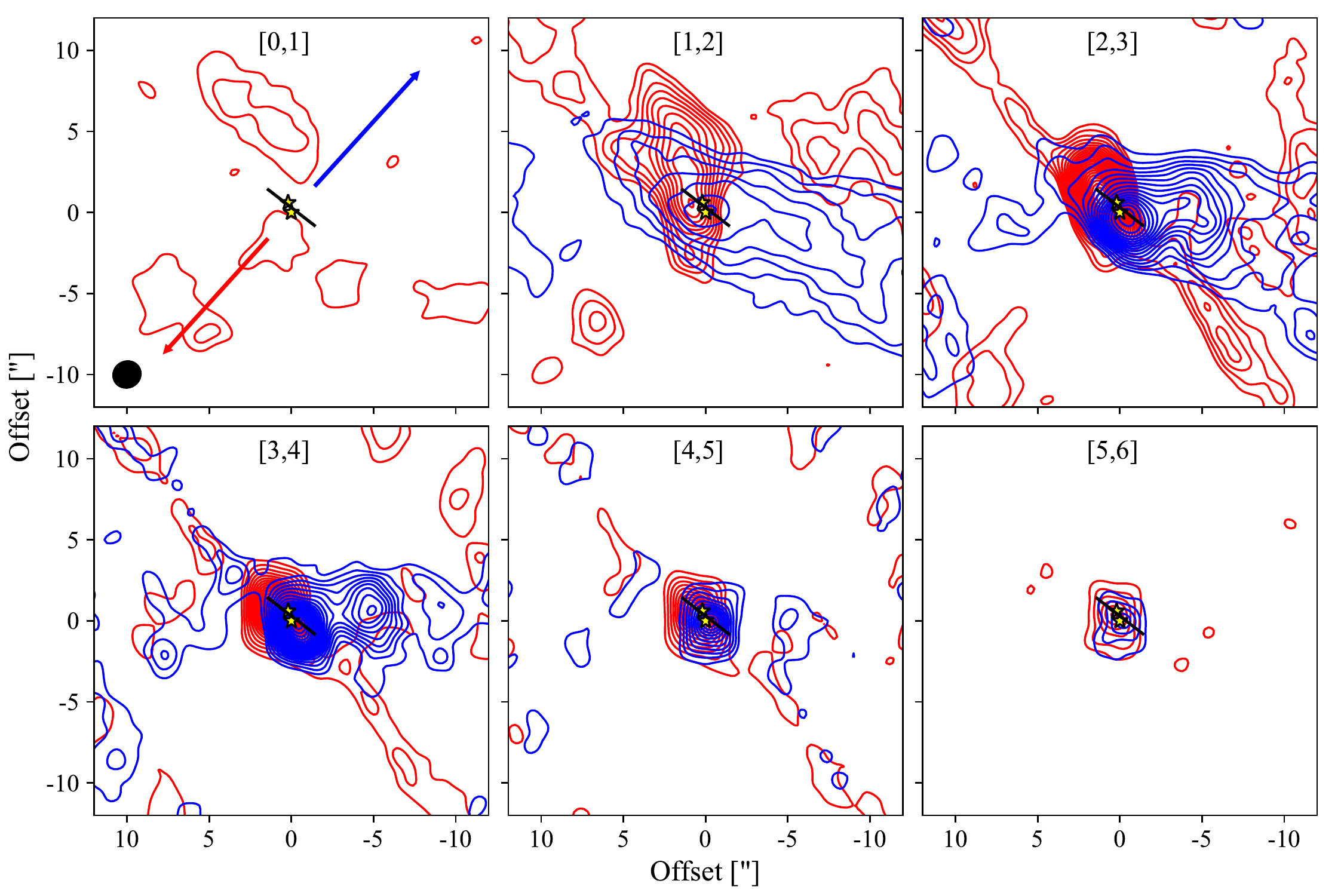}
      \caption[]{\label{fig:CO_cont}
      CO emission above 5$\sigma$ ($\sigma$~=~0.04 Jy~beam$^{-1}$~km~s$^{-1}$). The contours start at 5$\sigma$ and follow a step of 5$\sigma$. The numbers in brackets indicate the velocity interval in units of km~s$^{-1}$ and the extent of the circumbinary disc is represented by the black solid segment. The yellow stars show the position of the sources. The synthesised beam and the outflow direction from \cite{Bontemps1996} are represented in the upper-left panel with a black filled ellipse and blue and red arrows, respectively.
   }
\end{figure*}

\begin{figure*}[htb]
   \centering
      \includegraphics[width=.8\textwidth]{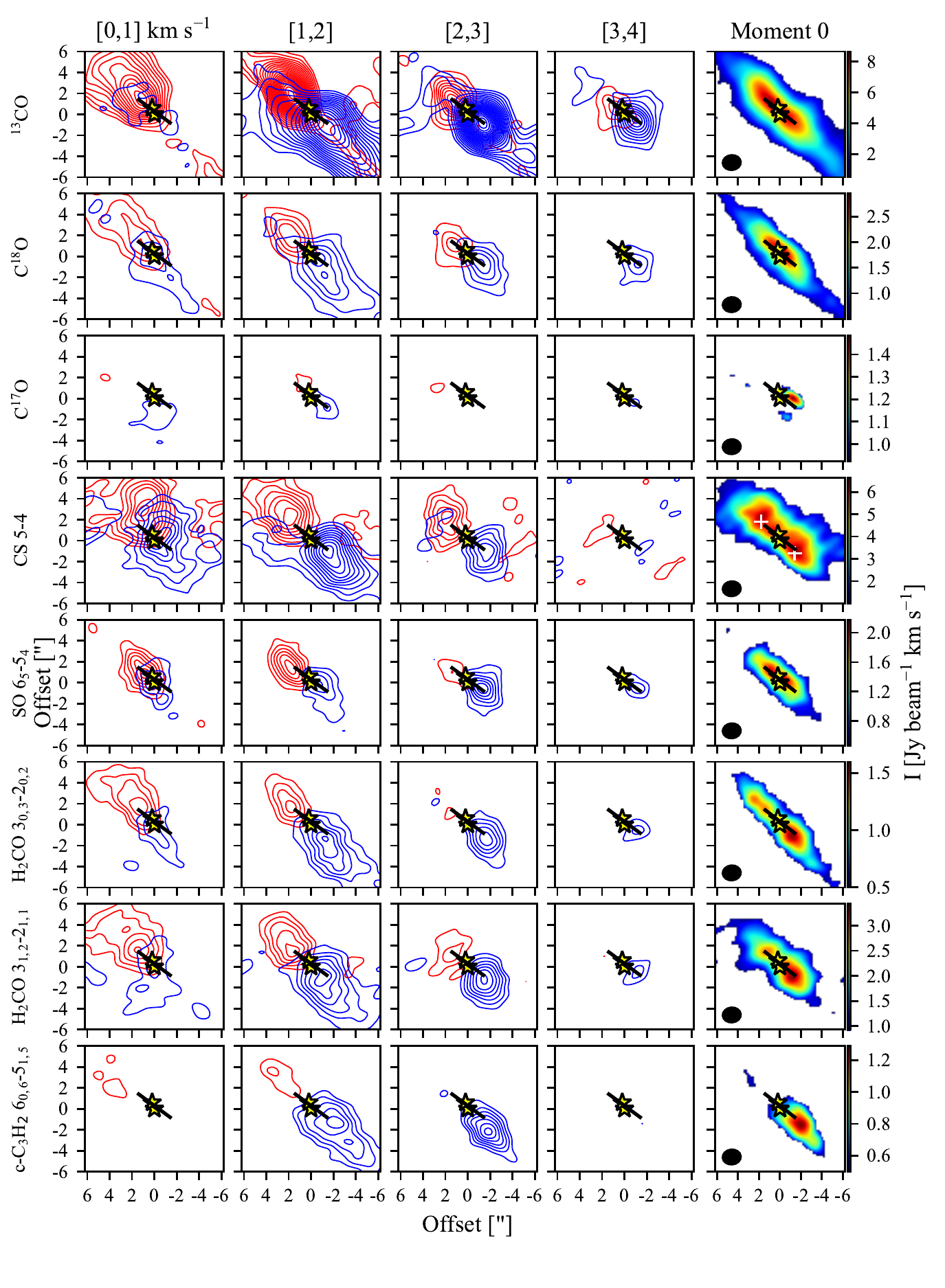}
      \caption[]{\label{fig:ALL}
      Emission of CO isotopologues and other bright molecular transitions. Channel maps consist of velocity ranges of 1~km~s$^{-1}$ and moment 0 maps are integrated over a velocity range of 8~km~s$^{-1}$. The contours start at 5$\sigma$ and follow a step of 5$\sigma$ for $^{13}$CO, C$^{18}$O, and CS (the remaining follow a step of 3$\sigma$). The two white crosses in the CS moment 0 map indicate the position from which the spectra from Figs.~\ref{fig:spec_all} and \ref{fig:spec_CN} are extracted. The yellow stars show the position of the sources and the solid segment represents the extent of the circumbinary disc. The synthesised beam is represented by a solid black ellipse in the right panels. 
   }
\end{figure*}

\subsection{Molecular transitions}

The observations covered multiple molecular transitions, such as CO isotopologues, S-bearing species, deuterated species, and carbon-chain molecules. The detected transitions are listed in Table~\ref{table:detected_molecules} with their spectroscopic parameters, while the lines without detection above a 3$\sigma$ level are listed in Table~\ref{table:non_detected_molecules} in the Appendix.

\subsubsection{CO isotopologues}

The CO emission is shown in Fig.~\ref{fig:CO_cont} for different velocity ranges. Most of the CO emission is expected to be extended and thus, filtered out by the interferometer due to the lack of short spacings. By comparing APEX data \citep{Lindberg2017} with convolved SMA C$^{18}$O emission (see Fig.~\ref{fig:APEX_SMA} and Table~\ref{table:APEX} in the Appendix), the recovered C$^{18}$O flux with the SMA is less than 1$\%$. Since CO is the more abundant isotopologue, and its emission is expected to be more extended than C$^{18}$O, the filtering out of emission is more significant for CO and thus, the recovered CO flux is expected to be much lower than 1$\%$. This implies that the SMA observations are probing the deepest and denser regions, and the detected CO emission in Fig.~\ref{fig:CO_cont} traces the more compact structures. The emission is centred at the position of the system, probing outer-envelope material, and knot-like structures are detected in red-shifted emission between 1 and 2~km~s$^{-1}$ and blue-shifted emission between 3 and 4~km~s$^{-1}$, which are consistent with the outflow direction found by \cite{Bontemps1996} (see upper-left panel of Fig.~\ref{fig:CO_cont}). 

Emission from $^{13}$CO, C$^{18}$O, and C$^{17}$O is shown in Fig.~\ref{fig:ALL}. $^{13}$CO presents a flatted shape that follows the circumbinary disc structure and extends beyond it. C$^{18}$O follows a similar distribution as $^{13}$CO, and C$^{17}$O stands out towards the southern part of the disc. Another C$^{17}$O transition (\textit{J}=3$-$2) is analysed by \cite{Artur2018}, where the emission traces the Keplerian circumbinary disc, therefore, the differences seen between C$^{17}$O \textit{J}=3$-$2 and C$^{17}$O \textit{J}=2$-$1 may be related to the sensitivity of the observations. A rotational profile, perpendicular to the outflow direction, is clearly seen for $^{13}$CO and C$^{18}$O for velocities beyond $\pm$2~km~s$^{-1}$.

\subsubsection{Sulphur-bearing species} 

Emission of CS and the brightest SO transition (6$_{5}$$-$5$_{4}$) is also shown in Fig.~\ref{fig:ALL}. The high dipole moment of CS makes this molecule a very good high-density tracer \citep[e.g.][]{vanderTak2000}. Its emission shows a large extent ($\sim$10$\arcsec$) and peaks at the edges of the circumbinary disc, showing a rotational profile around the binary system. SO shows a more compact distribution than CS, tracing the circumbinay disc material. SO 6$_{5}$$-$5$_{4}$ is slightly brighter than SO 5$_{5}$$-$4$_{4}$ (see Fig.~\ref{fig:SO} in the Appendix) and a third SO transition (1$_{2}$$-$2$_{1}$) was also targeted, however, no emission is detected above a 3$\sigma$ level. This may be related with its low \textit{A$_{ij}$}, which is two orders of magnitude lower than the value from the detected SO transitions (see Tables~\ref{table:detected_molecules} and \ref{table:non_detected_molecules} in the Appendix).

Less abundant CS isotopologues, such as C$^{34}$S and $^{13}$CS, are shown in Fig.~\ref{fig:CS_isotop_H2CO} in the Appendix, where C$^{34}$S is more than a factor of two brighter than $^{13}$CS, in agreement with their relative abundances with respect to CS \citep[$^{32}$S/$^{34}$S=22 and $^{12}$C/$^{13}$C=69;][]{Wilson1999}. C$^{34}$S follows the CS emission, where both species peak at the edges of the circumbinary disc. $^{13}$CS, on the other hand, presents isolated peaks that lack a clear correlation with the main isotopologue. The emission does, however, have a low signal-to-noise level ($\leq$4$\sigma$).

\subsubsection{H$_{2}$CO} 

The observations include six H$_{2}$CO transitions, associated with \textit{E$_\mathrm{u}$} values from 21 to 280~K. The transitions with the highest \textit{E$_\mathrm{u}$} (174 and 280~K) are not detected (see Table~\ref{table:non_detected_molecules}) and the strongest lines are o-H$_{2}$CO 3$_{1,2}$$-$2$_{1,1}$ (\textit{E$_\mathrm{u}$} = 33~K) and p-H$_{2}$CO 3$_{0,3}$$-$2$_{0,2}$ (\textit{E$_\mathrm{u}$} = 21~K), shown in Fig.~\ref{fig:ALL}. The emission extends beyond the circumbinary disc structure, peaks towards the edges of the disc, and the intensity decreases towards the positions of the protostars. This is similar to what is seen for CS, however, the H$_{2}$CO emission stands out towards the southern part of the circumbinary disc and the peak is slightly south from the disc. It is interesting to notice that the southern part of the disc is related with blue-shifted emission, while the blue component of the outflow is seen towards the north-west.

Two H$_{2}$CO transitions (3$_{2,2}$$-$2$_{2,1}$ and 3$_{2,1}$$-$2$_{2,0}$) show weak emission and are presented in Fig.~\ref{fig:CS_isotop_H2CO} in the Appendix. Their emission do not follow the distribution from the brightest H$_{2}$CO transition (3$_{1,2}$$-$2$_{1,1}$), however, both of them are associated with higher \textit{E$_\mathrm{u}$} values of 68~K. As previously shown \citep[e.g.][]{Mangum1993}, and further discussed in Section~4.1, the intensity ratios of these specific H$_{2}$CO transitions are sensitive tracers of the gas kinetic temperature and density.

\subsubsection{Carbon-chain molecules} 

The C$_{3}$H$_{2}$ molecule has two isomeric forms: cyclic (c-C$_{3}$H$_{2}$) and linear (l-C$_{3}$H$_{2}$). Although standard astrochemical models predict a cyclic-to-linear C$_{3}$H$_{2}$ ratio of $\sim$1, observations show higher values ranging from 3 to 70 \citep{Sipila2016}. Our observations targeted nine c-C$_{3}$H$_{2}$ transitions, with \textit{E$_\mathrm{u}$} between 19 and 87~K, and three l-C$_{3}$H$_{2}$ transitions, with \textit{E$_\mathrm{u}$} between 66 and 80~K (see Tables~\ref{table:detected_molecules} and \ref{table:non_detected_molecules}). However, the linear isomer is not detected. Figure~\ref{fig:ALL} shows the brightest c-C$_{3}$H$_{2}$ transition, where the emission stands out between $-$3 and $-$1~km~s$^{-1}$ and peaks south from the circumbinary disc. Other three c-C$_{3}$H$_{2}$ lines present weaker emission and are shown in Fig.~\ref{fig:C3H2_HC3N} in the Appendix. 

The brightest c-C$_{3}$H$_{2}$ transition (6$_{0,6}$$-$5$_{1,5}$) was also detected with APEX by \cite{Lindberg2017} (see Appendix), and more than 92$\%$ of the flux is filtered out by the interferometer, suggesting that the c-C$_{3}$H$_{2}$ distribution is much more extended than what is seen in Fig.~\ref{fig:ALL}. In addition, c-C$_{3}$H$_{2}$ is the only transition in the APEX data that peaks at 2.9~km~s$^{-1}$, and not at 4.2~km~s$^{-1}$ like the rest of the detected molecular transitions (DCO$^{+}$, H$_{2}$CO, C$^{18}$O, and SO). The velocity of 2.9~km~s$^{-1}$ is consistent with the blue-shifted material detected in the SMA observations, where the c-C$_{3}$H$_{2}$ emission stands out between $-$3 and $-$1~km~s$^{-1}$ with respect to the binary system velocity. 

Three HC$_{3}$N transitions are detected and shown in Fig.~\ref{fig:C3H2_HC3N} in the Appendix. The emission peaks south of the circumbinary disc, which is consistent with the c-C$_{3}$H$_{2}$ emission, suggesting that this region is rich in carbon-chain molecules. On the other hand, the HC$_{3}$N \textit{J}=24$-$23 transition was observed with APEX, but not detected, therefore, the emission may arise from a compact component (i.e. the emission is suffering from beam dilution).

\subsubsection{CN}

\begin{figure*}[t]
   \centering
      \includegraphics[width=.7\textwidth]{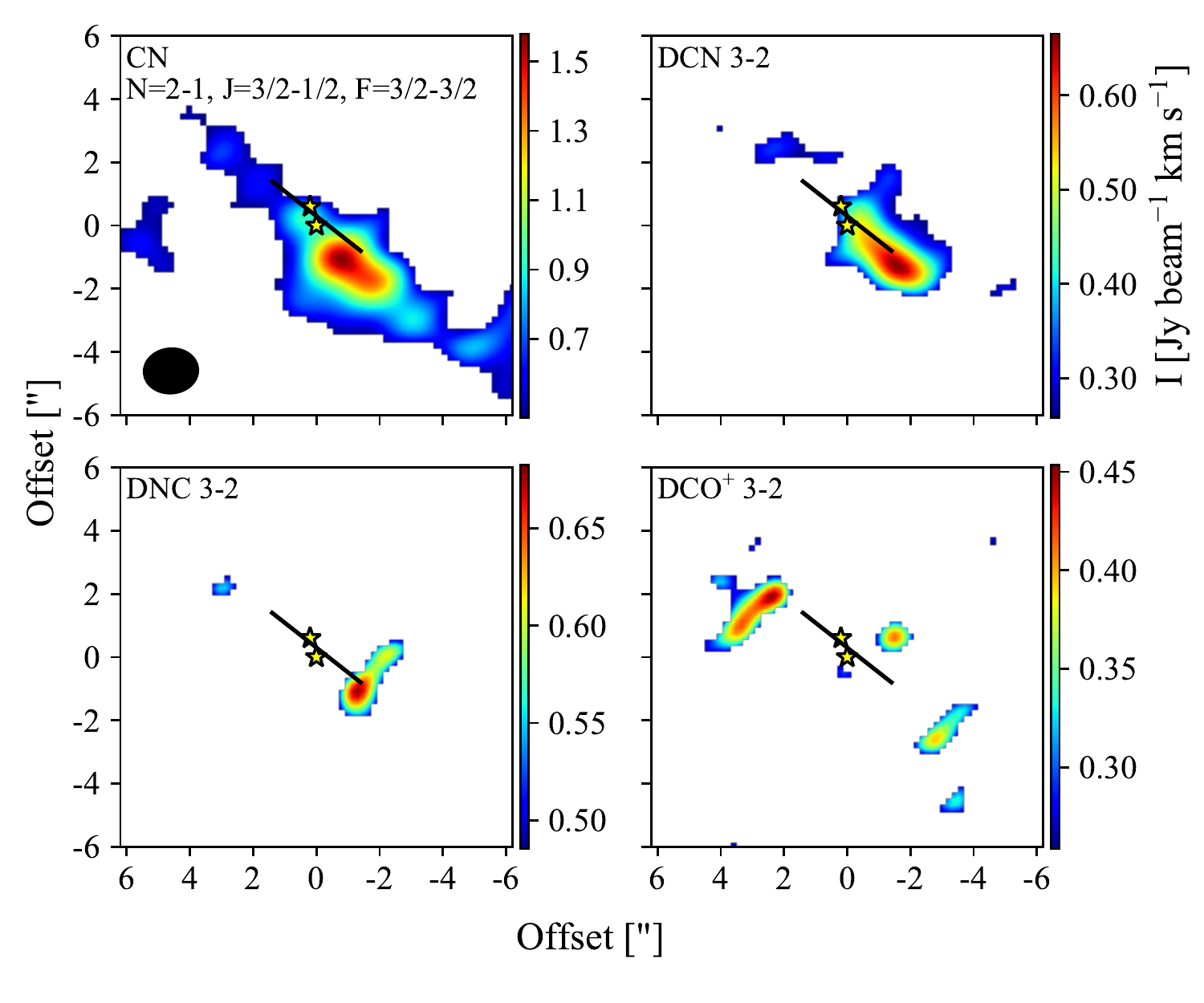}
      \caption[]{\label{fig:D_CN}
      CN and deuterated species, DCN, DNC, and DCO$^{+}$, moment 0 maps integrated over a velocity range of 6~km~s$^{-1}$. The yellow stars show the position of the sources and the solid segment represents the extent of the circumbinary disc. The synthesised beam is represented by a black filled ellipse in the upper-left panel.  
   }
\end{figure*}

CN is know as a photon-dominated region (or photodissociation region, PDR) tracer, since its abundance increases when HCN photo-dissociates into CN due to ultraviolet (UV) radiation from the star or the interstellar radiation field \citep{Willacy2000, Aikawa2001, vanZadelhoff2003}. Although gas-phase CN is expected to be abundant towards protostars, it is often challenging to observe with interferometers at millimeter wavelengths as it typically is extended and filtered out \citep[e.g.][]{Jorgensen2011}. Towards one Class 0 source, L483-mm, CN was found to probe material in the boundary between the bulk protostellar envelope and its outflow cavity \citep{Jorgensen2004d}. CN has also been detected towards a number of protostellar discs around Class II sources \citep[e.g.][]{Dutrey1997, Qi2001, Thi2004, Oberg2011b}, where it is found to be a good tracer of the disc surfaces \citep[e.g.][]{vanDishoeck2006}. 

Toward IRS67, ten of the twelve hyperfine transitions from CN 2$-$1 (see Table~\ref{table:detected_molecules}) are detected. The emission of one of these hyperfine transitions is shown in Fig.~\ref{fig:D_CN}. The brightest CN lines are blended with other CN hyperfine transitions and their moment 0 maps are shown in Fig.~\ref{fig:CN_combination} in the Appendix. The CN emission stands out towards the southern region and peaks slightly offset from the southern edge of the circumbinary disc, which is consistent with the H$_{2}$CO peak (see Fig.~\ref{fig:ALL}).

\subsubsection{Deuterated species} 

Deuterated species are good tracers of cold regions, in particular, the outer disc midplane where the temperature is $\leq$~30~K \citep[e.g.][]{vanDishoeck2006, Jorgensen2011, Murillo2015, Oberg2015, Aikawa2018}. The ion DCO$^{+}$ is apparently formed in the gas phase, considering its short destruction timescale and is expected to be abundant in regions with temperature between 19 and 21~K \citep{Aikawa2018}. On the other hand, neutral species as DCN can form in interstellar ice and later desorb in the disc, thus tracing regions with higher temperatures than DCO$^{+}$ \citep[$\geq$~30~K;][]{Jorgensen2004c, Aikawa2018}. Emission of DCN, DNC, and DCO$^{+}$ is detected and shown in Fig.~\ref{fig:D_CN}. Both DCN and DNC peak towards the southern edge of the protobinary disc, which is consistent with the CN peak, while DCO$^{+}$ stands out close to the northern edge of the disc and no emission is detected where DCN and DNC peak.

\subsubsection{Spectra}

\begin{figure*}[t]
   \centering
      \includegraphics[width=.98\textwidth]{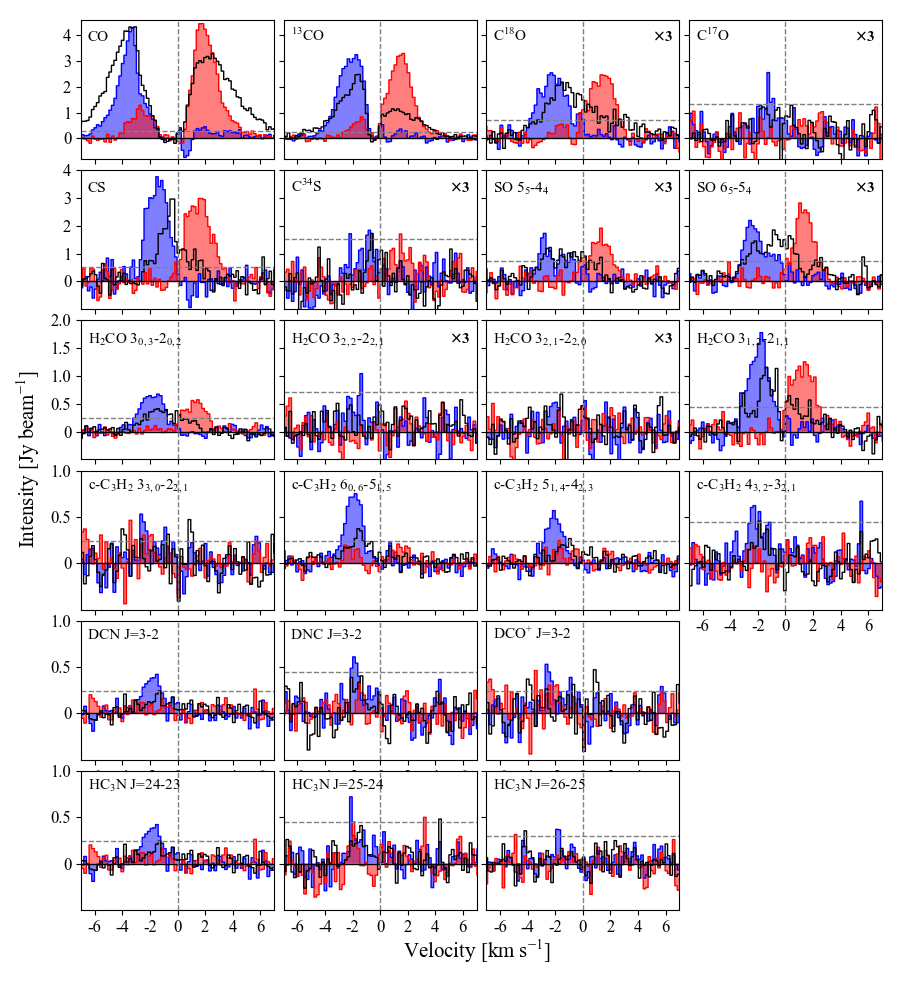}
      \caption[]{\label{fig:spec_all}
      Spectra of the brightest molecular transitions (with the exception of CN) taken at three different positions: the geometric centre (black), the southern edge (blue), and the northern edge of the circumbinary disc (red; see the white crosses in the moment 0 map of CS in Fig.~\ref{fig:ALL}). The zero velocity value (dashed grey vertical line) corresponds to a system velocity of 4.2~km~s$^{-1}$. The dashed grey horizontal line shows the value of 3$\sigma$ (see Table~\ref{table:observations}). Some of the spectra are multiplied by a factor of three, as indicated in the top right corner.
         }
\end{figure*}

\begin{figure*}[t]
   \centering
      \includegraphics[width=.98\textwidth]{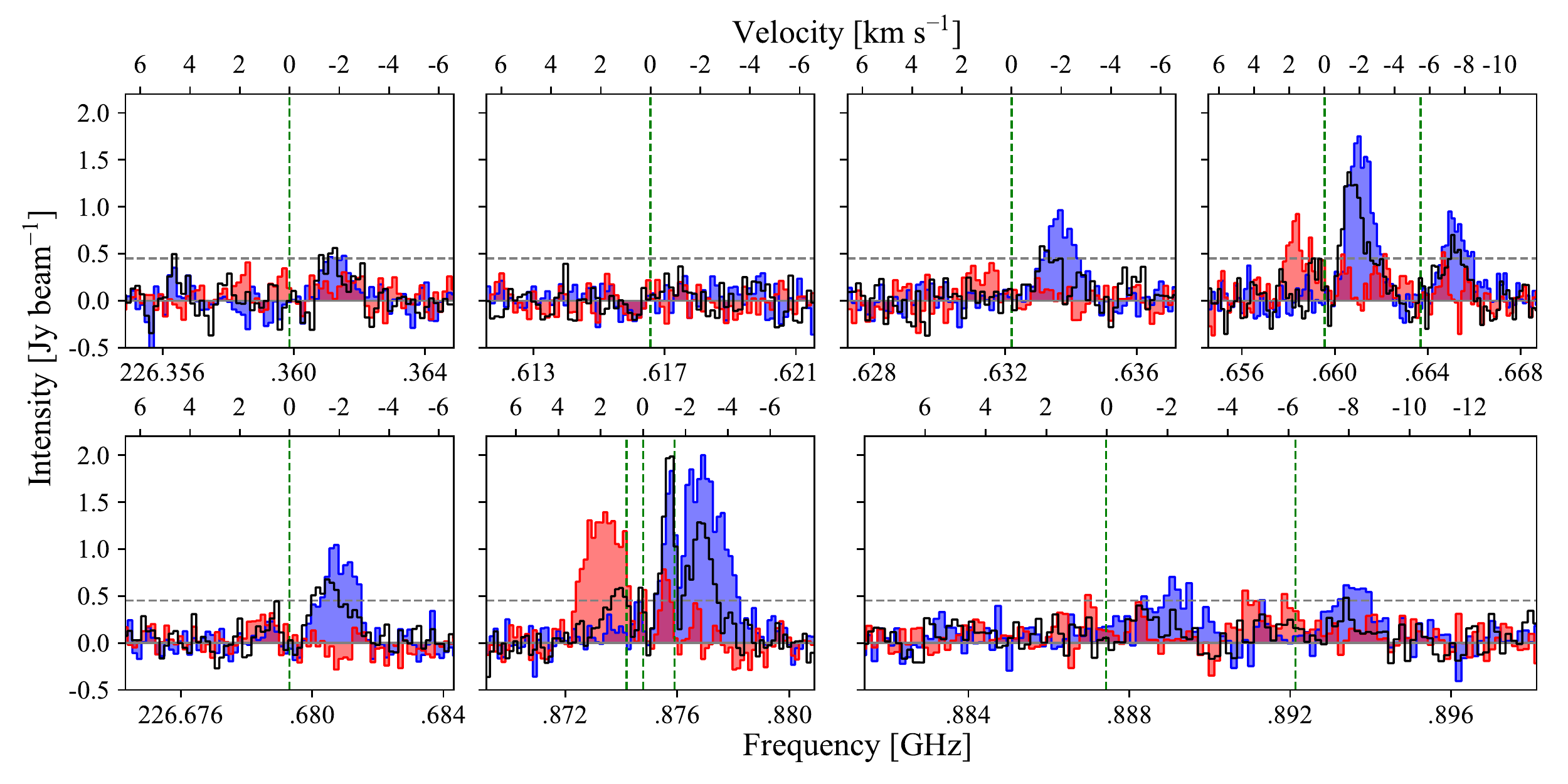}
      \caption[]{\label{fig:spec_CN}
      CN spectra taken at three different positions: the geometric centre (black), the southern edge (blue), and the northern edge of the circumbinary disc (red; see white crosses in the moment 0 map of CS in Fig.~\ref{fig:ALL}). The dashed green vertical lines represent the rest frequency of the CN hyperfine transitions (see Table~\ref{table:detected_molecules}), while the dashed grey horizontal line shows the value of 3$\sigma$. 
   }
\end{figure*}

Figure~\ref{fig:spec_all} shows the spectra of the species presented in Figs.~\ref{fig:ALL} and \ref{fig:D_CN}, towards three different positions: the geometric centre and the blue- and red-shifted peaks from the CS emission, located at 2$\arcsec$ and 2.4$\arcsec$ from the geometric centre, respectively (see white crosses in the moment 0 map of CS in Fig.~\ref{fig:ALL}). CO and $^{13}$CO show a clear absorption future at the system velocity (4.2~km~s$^{-1}$), mainly due to the interferometric filtering-out of emission from large scales. C$^{18}$O, CS, SO, and the two brightest H$_{2}$CO transitions present similar spectral features, with prominent emission between $-$3 and 3~km~s$^{-1}$ and a considerably symmetry between the blue- and red-shifted components. These components are tracing the edges of the circumbinary disc, peak around $\pm$2~km~s$^{-1}$ and are consistent with a rotational profile. This is in agreement with the results from \cite{Artur2018}, where velocities below $\pm$2.5~km~s$^{-1}$ are expected at a distance of 2$\arcsec$ from the geometric centre and beyond. On the other hand, c-C$_{3}$H$_{2}$, DCN, DNC, and HC$_{3}$N show strong emission towards the southern edge of the circumstellar disc (blue spectra), with a peak around $-$2~km~s$^{-1}$, tracing a different component than the CO isotopologues, CS, SO and H$_{2}$CO.

Since the observations include multiple CN hyperfine transitions, their spectra are shown in Fig.~\ref{fig:spec_CN}, covering a frequency range from 226.356 to 226.896~GHz. The spectra of isolated lines show an absorption feature at the system velocity, consistent with filtering-out of emission from large scales. This agrees with the results from \cite{Jorgensen2011}, where the CN emission may be tracing the outer envelope towards IRAS 16293-2422 and is filtered out by the SMA. Nevertheless, towards IRS 67 the CN emission peaks at the southern edge of the circumbinary disc, with velocities around $-$2~km~s$^{-1}$. The CN spectra resemble that of  c-C$_{3}$H$_{2}$ (Fig.~\ref{fig:spec_all}), suggesting that the two species trace the same material.

\section{Discussion}

\subsection{Structure of IRS 67}

\begin{figure}[t]
   \centering
      \includegraphics[width=.48\textwidth]{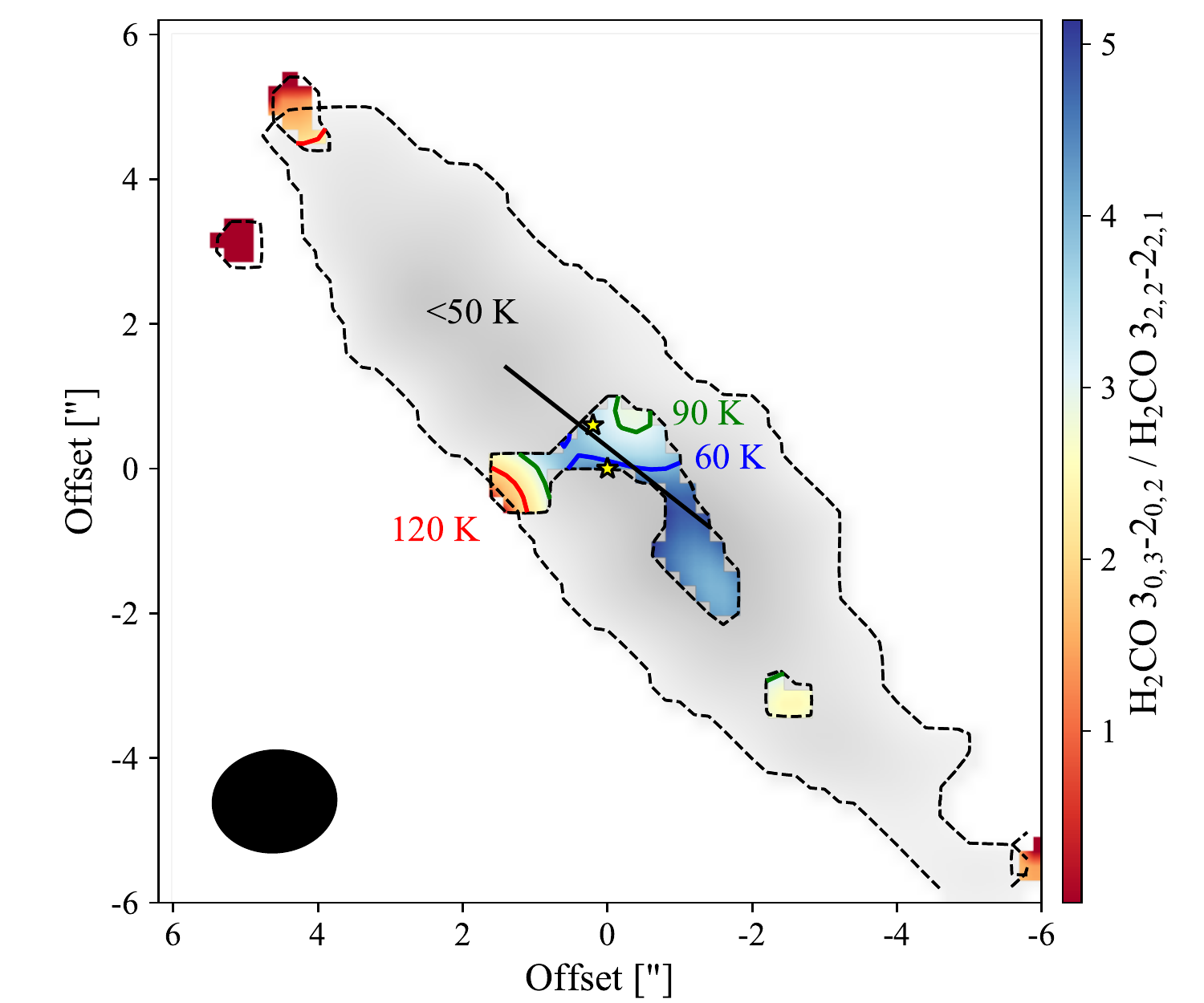}
      \caption[]{\label{fig:H2CO_ratio}
      Gas temperature estimation from the ratio between o-H$_{2}$CO 3$_{0,3}$$-$2$_{0,2}$ and o-H$_{2}$CO 3$_{2,2}$$-$2$_{2,1}$, following the results from \cite{Mangum1993}. Specific values of 60, 90, and 120~K are shown by blue, green, and red contours, respectively. The grey region, where o-H$_{2}$CO 3$_{2,2}$$-$2$_{2,1}$ is not detected, indicates temperatures below 50~K. The yellow stars show the position of the sources and the solid segment represents the extent of the circumbinary disc. The synthesised beam is represented by a black filled ellipse. 
   }
\end{figure}

The ratio between H$_{2}$CO 3$_{0,3}$$-$2$_{0,2}$ and H$_{2}$CO 3$_{2,2}$$-$2$_{2,1}$ has previously been shown to be a good tracer of the gas temperature \citep{Mangum1993}. Figure~\ref{fig:H2CO_ratio} shows the estimated gas temperature for a H$_{2}$ number density $\geq$ 10$^{8}$~cm$^{-3}$, following those results. The highest temperatures ($\geq$90~K) are seen towards the outflow direction, while the southern edge of the circumbinary disc is associated with temperatures between 50 and 60~K. This is the region where the peak of emission is seen for DCN, DNC, CN, c-C$_{3}$H$_{2}$, and HC$_{3}$N, suggesting that these species are tracing high densities (\textit{n$_\mathrm{H_{2}}$} $\geq$ 10$^{8}$~cm$^{-3}$) and lukewarm temperatures (50$-$60~K). The non-detection of H$_{2}$CO 3$_{2,2}$$-$2$_{2,1}$ where H$_{2}$CO 3$_{0,3}$$-$2$_{0,2}$ is detected sets an upper limit of 50~K for the gas temperature (the grey region in Fig.~\ref{fig:H2CO_ratio}). 

CN and c-C$_{3}$H$_{2}$ are commonly used as PDR tracers, since their emission shows a strong dependence on the UV radiation \citep[e.g.][]{Dutrey1997, vanDishoeck2006, Oberg2011b, Bergin2016, Murillo2018}. The circumbinary disc associated with IRS 67 may be strongly illuminated by UV radiation from the binary system. Because discs are normally flared, is it also likely that the southern region of the circumbinary disc is associated with a PDR, in particular, the surface layer of the circumbinary disc. Other common PDRs originate along the outflow cavity \citep[e.g.][]{Murillo2018}, however, this is unlikely for IRS 67 since the blue-shifted outflow component is seen towards the north-west, and blue-shifted emission from c-C$_{3}$H$_{2}$ is detected towards the south-east. The PDR originating at the surface layer of the circumbinary disc is also consistent with the temperature structure estimated from the H$_{2}$CO 3$_{0,3}$$-$2$_{0,2}$/3$_{2,2}$$-$2$_{2,1}$ ratio (see Fig.~\ref{fig:H2CO_ratio}). 

The northern part of the circumbinary disc seems to be related with colder regions than the southern part (see Fig.~\ref{fig:H2CO_ratio}). DCO$^{+}$ is expected to trace cold regions \citep[$\leq$30~K;][]{Jorgensen2011, Aikawa2018} and its emission peaks towards the north-east of the circumbinary disc (Fig.~\ref{fig:D_CN}), showing an anti-correlation with the PDR tracers (mainly CN and c-C$_{3}$H$_{2}$). The DCO$^{+}$ emission seems to be associated with cold regions from the inner envelope at small scales ($\leq$1000~AU), however, the same DCO$^{+}$ transition was detected with APEX (see Fig.~\ref{fig:APEX_SMA} in the Appendix), suggesting that more than 96$\%$ of the emission is filtered out by the interferometer. Therefore, DCO$^{+}$ is expected to be present also at large scales \citep[e.g.][]{Jorgensen2011, Murillo2018}. 

The system IRS 67 is particularly rich in molecular lines when it is compared with other single Class I sources \citep{Artur2019}, possibly due to the mass content and extent of the circumbinary disc. One of the species that is only detected towards IRS 67, among other 11 Class I sources, is c-C$_{3}$H$_{2}$, where two transitions (5$_{5,1}$$-$4$_{4,0}$ and 5$_{5,0}$$-$4$_{4,1}$) are seen \citep{Artur2018} and the emission is consistent with the c-C$_{3}$H$_{2}$ lines analysed in this work. c-C$_{3}$H$_{2}$ may, therefore, be related with PDRs (like outflow cavities and disc surface layers) but also with high gas column densities expected towards Class 0 sources and the binary system IRS 67. This is in agreement with the non-detection of c-C$_{3}$H$_{2}$ towards a sample of 12 Class II discs \citep{Oberg2010, Oberg2011b}.

\subsection{Non-detections}

Together with the detection of multiple molecular transitions, the non-detections may provide some clues about the physical parameters of the region. The main non-detections at the 3$\sigma$ level are N$_{2}$D$^{+}$, the linear isomer l-C$_{3}$H$_{2}$, SiO, and CH$_{3}$OH. 

N$_{2}$D$^{+}$ is a good tracer of cold regions ($\leq$30~K), where CO freezes out, and usually correlates with DCO$^{+}$ \citep[e.g.][]{Jorgensen2011, Aikawa2018, Murillo2018}. Since most of the DCO$^{+}$ is filtered out by the interferometer and its emission is expected to be present at large scales, N$_{2}$D$^{+}$ may also be present at large scales. Therefore, the non-detection of N$_{2}$D$^{+}$ may be related to the lack of short-spacings in the observations.

The non-detection of the linear C$_{3}$H$_{2}$ isomer may be related with the high cyclic-to-linear ratio or to an excitation effect, or a combination of both. The cyclic-to-linear ratio has been proven to differ from the statistical value of one for a variety of objects, ranging from 3 to 70 \citep{Sipila2016}. On the other hand, the three l-C$_{3}$H$_{2}$ transitions covered in the observations have \textit{E$_\mathrm{u}$} values between 66 and 80~K, while the brightest c-C$_{3}$H$_{2}$ line (see Fig.~\ref{fig:ALL}) has \textit{E$_\mathrm{u}$}~=~39~K. In addition, c-C$_{3}$H$_{2}$ 7$_{2,6}$$-$7$_{1,7}$ (\textit{E$_\mathrm{u}$} = 61~K) shows weak emission and two c-C$_{3}$H$_{2}$ lines with \textit{E$_\mathrm{u}$} = 87~K are not detected. This sets an upper limit of 60~K for the excitation temperature of the c-C$_{3}$H$_{2}$ emitting region. Therefore, a combination of both, high cyclic-to-linear ratio and an excitation effect, may explain the non detection of the linear C$_{3}$H$_{2}$ isomer.

Typically, SiO is seen as a good tracer of shocks as it is present in the gas-phase when refractory grains are destroyed \citep[e.g.][]{Bachiller1997, Codella2014, Gusdorf2008a, Gusdorf2008b}. Its non-detection towards IRS 67 is consistent with other studies which suggest that Class I sources are associated with less energetic outflows.

\subsection{Chemical differentiation around IRS 67}

\begin{figure}[t]
   \centering
      \includegraphics[width=.48\textwidth]{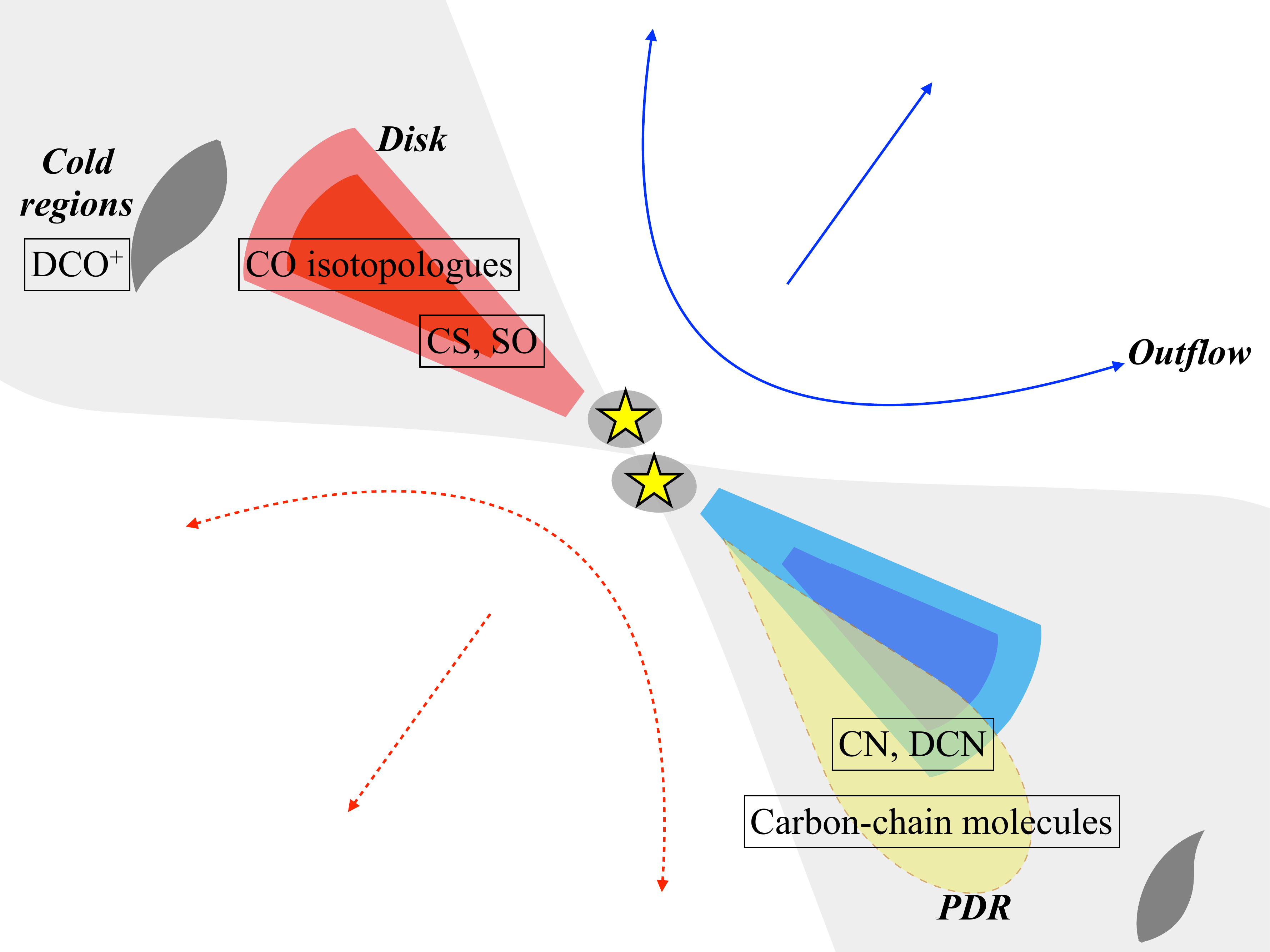}
      \caption[]{\label{fig:cartoon}
      Schematic representation of the environment towards IRS 67 where three main regions are distinguished. Cold regions traced by DCO$^{+}$, the disc structure proven by CO isotopologues and S-bearing species, and a PDR associated with the surface layers of the disc, traced by CN, DCN, and carbon-chain molecules. The outflow direction is taken from \cite{Bontemps1996}. 
   }
\end{figure}

Figure~\ref{fig:cartoon} shows a schematic representation of the environment associated with IRS 67, where three main regions are distinguished: (\textit{i}) a cold region (\textit{T}~$\leq$~30~K) beyond the extent of the circumbinary disc, traced by DCO$^{+}$, (\textit{ii}) the circumbinary disc traced by CO isotopologues and sulphur-bearing species, and (\textit{iii}) a PDR, likely the surface layer of the circumbinary disc reached by the UV radiation from the binary system, and traced by CN, DCN, and carbon-chain molecules. 

The chemistry towards IRS 67 shows some similarities with Class 0 sources, possibly due to the high-gas column density, and with Class II discs, where the UV radiation from the binary system may resemble the interaction between a typical T-Tauri star and the surface layers of its disc. Cold gas tracers (such as DCO$^{+}$ and DCN), potential grain chemistry products (such as H$_{2}$CO), and species associated with carbon-chain chemistry (such as c-C$_{3}$H$_{2}$ and HC$_{3}$N), are commonly detected towards Class 0 sources \citep[e.g.][]{Jorgensen2005b, Murillo2018} and are also detected towards IRS 67. However, while H$_{2}$CO and c-C$_{3}$H$_{2}$ are mainly seen at the outflow cavities towards Class 0 sources, towards IRS 67 they seem to trace the surface layers of the circumbinary disc. On the other hand, photochemistry products such as CN are abundant towards more evolved discs, associated with Class II sources \citep[e.g.][]{Oberg2010, Oberg2011b}: CN will originate in the outer, very low density part of the disc, which is completely exposed to UV radiation. The binary system IRS 67 may be associated with a stronger UV radiation field than single Class I sources, therefore, the CN emission towards IRS 67 appears to trace the same physical component as in Class II sources, that is, the surface layers of the disc exposed to UV radiation. Future observations of other Class I binary sources will verify whether the chemical richness of IRS 67 is common towards these stages, and higher angular resolutions will constrain the velocity profile and the dynamics of the detected transitions.

\section{Summary}

This work presents SMA observations of the Class I binary system IRS 67 in the Ophiuchus star forming region, with an angular resolution of 1$\farcs$7~$\times$1$\farcs$4 ($\sim$260~$\times$~210 AU). The continuum emission between 1.2 and 1.4~mm is analysed, together with multiple molecular transitions that trace different physics. The main results are summarised below.

\begin{itemize}
\item The continuum emission is consistent with previous studies \citep{Artur2018} and a power-law fitting results in a $\beta$-value of 0.8~$\pm$~0.3, suggesting that dust grains have grown to larger sizes than the ISM dust particles, or that the dust is optically thick. 
\item The detected molecular transitions are tracing three main regions: cold regions beyond the circumbinary disc extent, the circumbinary disc, and a PDR likely related with the surface layers of the disc. DCO$^{+}$ is tracing the cold regions, while the CO isotopologues and the sulphur-bearing species are probing the disc structure. In addition, CN, DCN, and carbon-chain molecules are tracing the PDR. 
\item The case of H$_{2}$CO is special, as it traces both the circumbinary disc and the PDR. The ratio between o-H$_{2}$CO 3$_{0,3}$$-$2$_{0,2}$ and o-H$_{2}$CO 3$_{2,2}$$-$2$_{2,1}$ has been shown to be a good indicator of the gas temperature, where the temperature map is consistent with the physical structure of IRS 67, that is, the warmer gas follows the outflow direction, lukewarm temperatures are associated with the PDR, and colder gas is related to the circumbinary disc.
\item IRS 67 shows chemical similarities with Class 0 sources, such as the detection of sulphur-bearing species and carbon-chain molecules, while PDR tracers, such as CN, are associated with Class II discs, where the UV radiation can reach the surface layers of the disc. IRS 67 is, therefore, a chemical link between these two stages.
\end{itemize}

This work shows the potential of the broad spectral coverage of the SMA, allowing us to detect and analyse multiple molecules and transitions from the same species. Similar observations of other Class I sources will provide more statistical results and highlight if IRS 67 is a particular chemically-rich system or it represents a general trend. In addition, higher angular resolution observations will constrain the dynamics of the gas and possibly resolve the individual circumstellar discs around each source, constraining the link between circumstellar and circumbinary discs.

\begin{acknowledgements}

We thank Johan Lindberg for sharing and discussing APEX data with us. This paper is based on data from the Submillimeter Array: the Submillimeter Array is a joint project between the Smithsonian Astrophysical Observatory and the Academia Sinica Institute of Astronomy and Astrophysics, and is funded by the Smithsonian Institution and the Academia Sinica. The group of JKJ acknowledges support from the European Research Council (ERC) under the European Union's Horizon 2020 research and innovation programme (grant agreement No 646908) through ERC Consolidator Grant ``S4F''. Research at the Centre for Star and Planet Formation is funded by the Danish National Research Foundation. 

\end{acknowledgements}

\begin{appendix}

\section{Other molecular transitions}

\begin{figure*}[t]
   \centering
      \includegraphics[width=.98\textwidth]{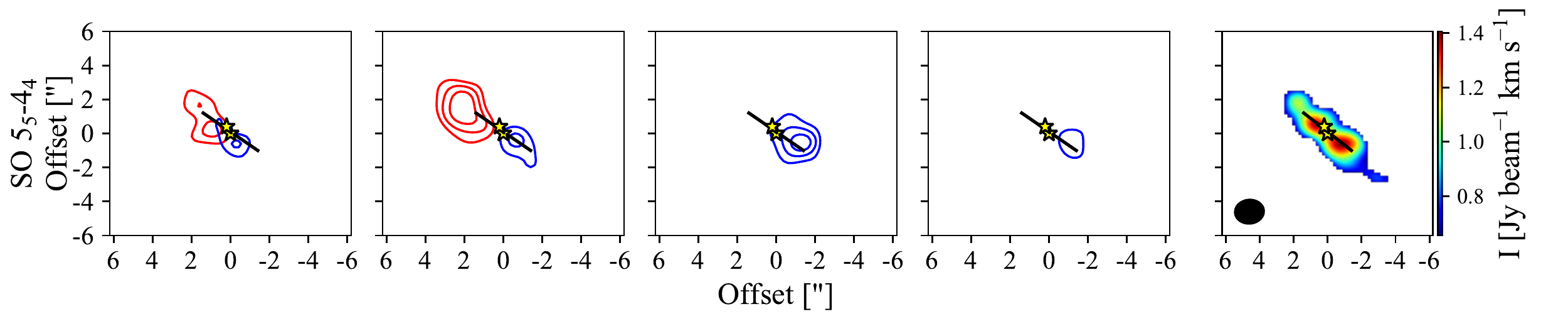}
      \caption[]{\label{fig:SO}
      Emission of SO 5$_{5}$$-$4$_{4}$. Channel maps consist of velocity ranges of 1~km~s$^{-1}$ and the moment 0 map is integrated over a velocity range of 8~km~s$^{-1}$. The contours start at 3$\sigma$ and follow a step of 3$\sigma$. The yellow stars show the position of the sources and the solid segment represents the extent of the circumbinary disc. The synthesised beam is represented in the right panel by a solid black ellipse. 
   }
\end{figure*}

\begin{figure*}[t]
   \centering
      \includegraphics[width=.48\textwidth]{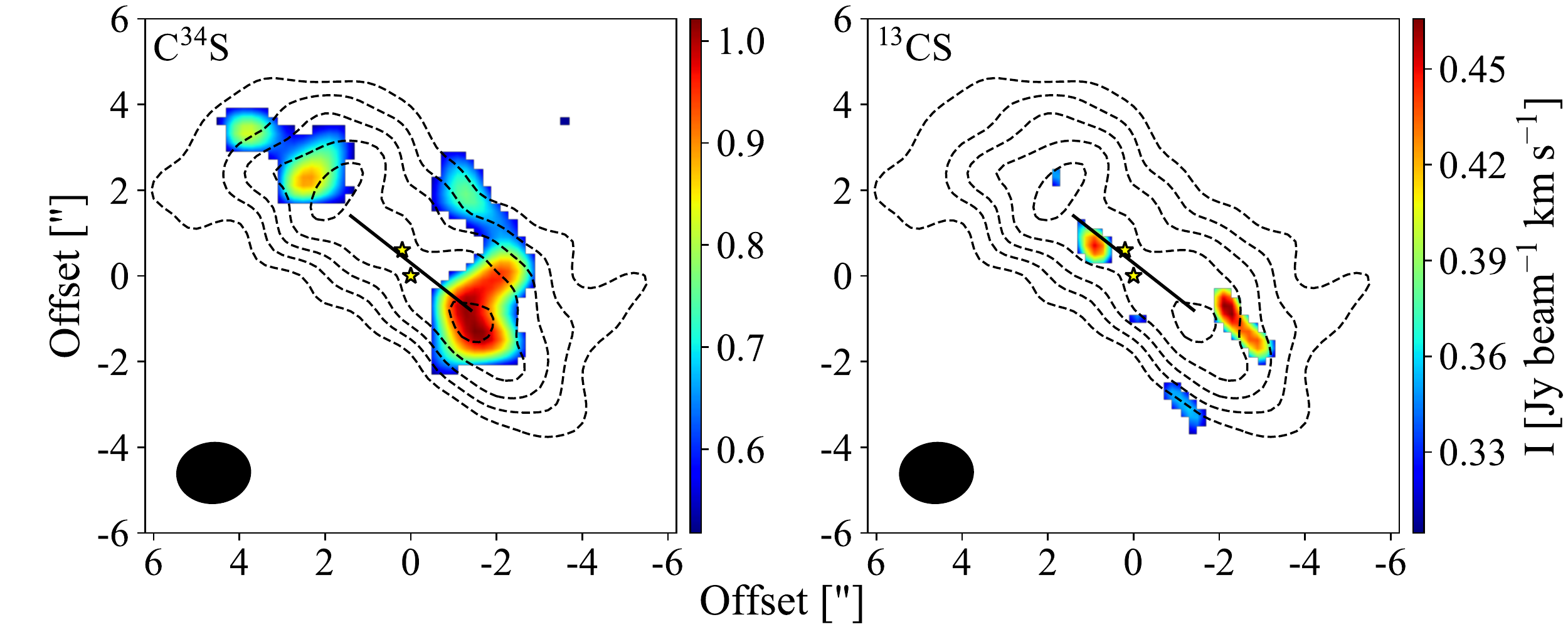}
      \includegraphics[width=.48\textwidth]{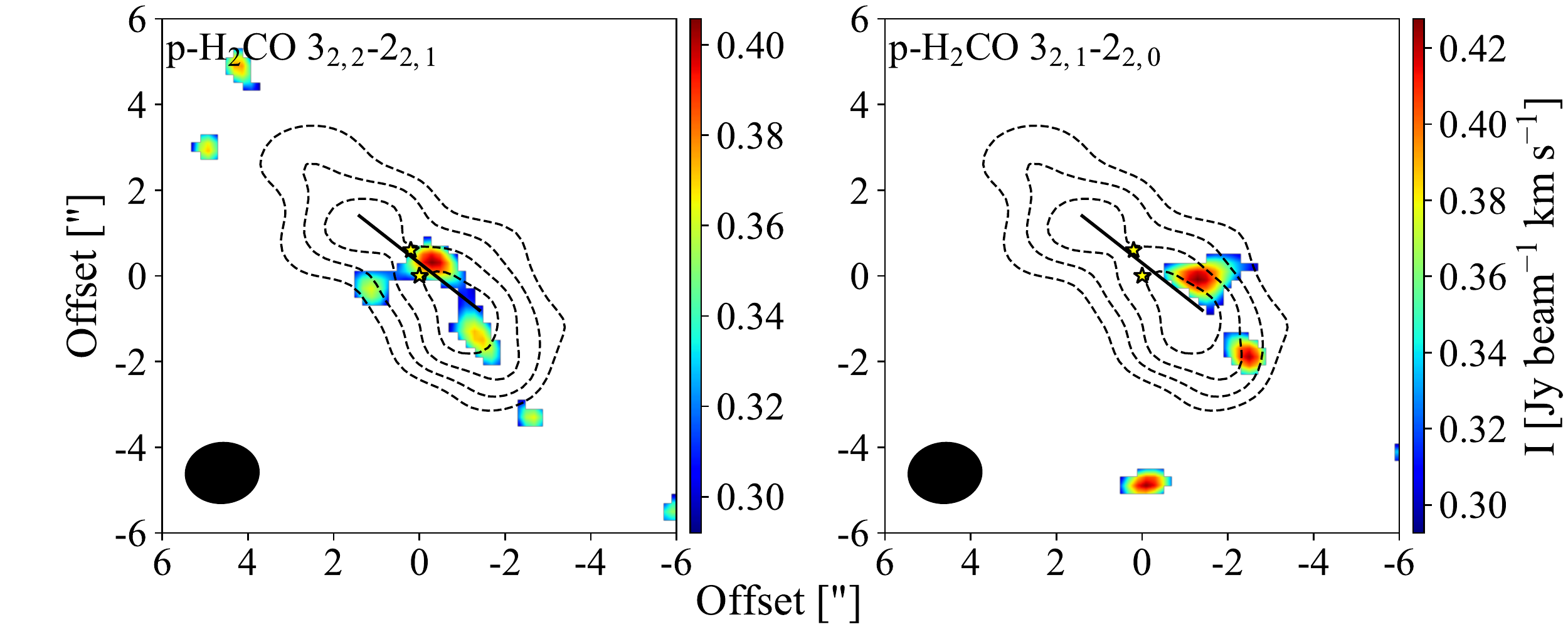}
      \caption[]{\label{fig:CS_isotop_H2CO}
      C$^{34}$S, $^{13}$CS, and H$_{2}$CO (\textit{E$_\mathrm{u}$} = 68~K) moment 0 maps (colour scale) integrated over a velocity range of 6~km~s$^{-1}$, overlapped with CS emission (contours in C$^{34}$S and $^{13}$CS maps) and H$_{2}$CO 3$_{1,2}$$-$2$_{1,1}$ (contours in H$_{2}$CO maps) from Fig.~\ref{fig:ALL}. The yellow stars show the position of the sources and the solid segment represents the extent of the circumbinary disc. The synthesised beam is represented by a black filled ellipse. 
   }
\end{figure*}

\begin{figure*}[t]
   \centering
      \includegraphics[width=.98\textwidth]{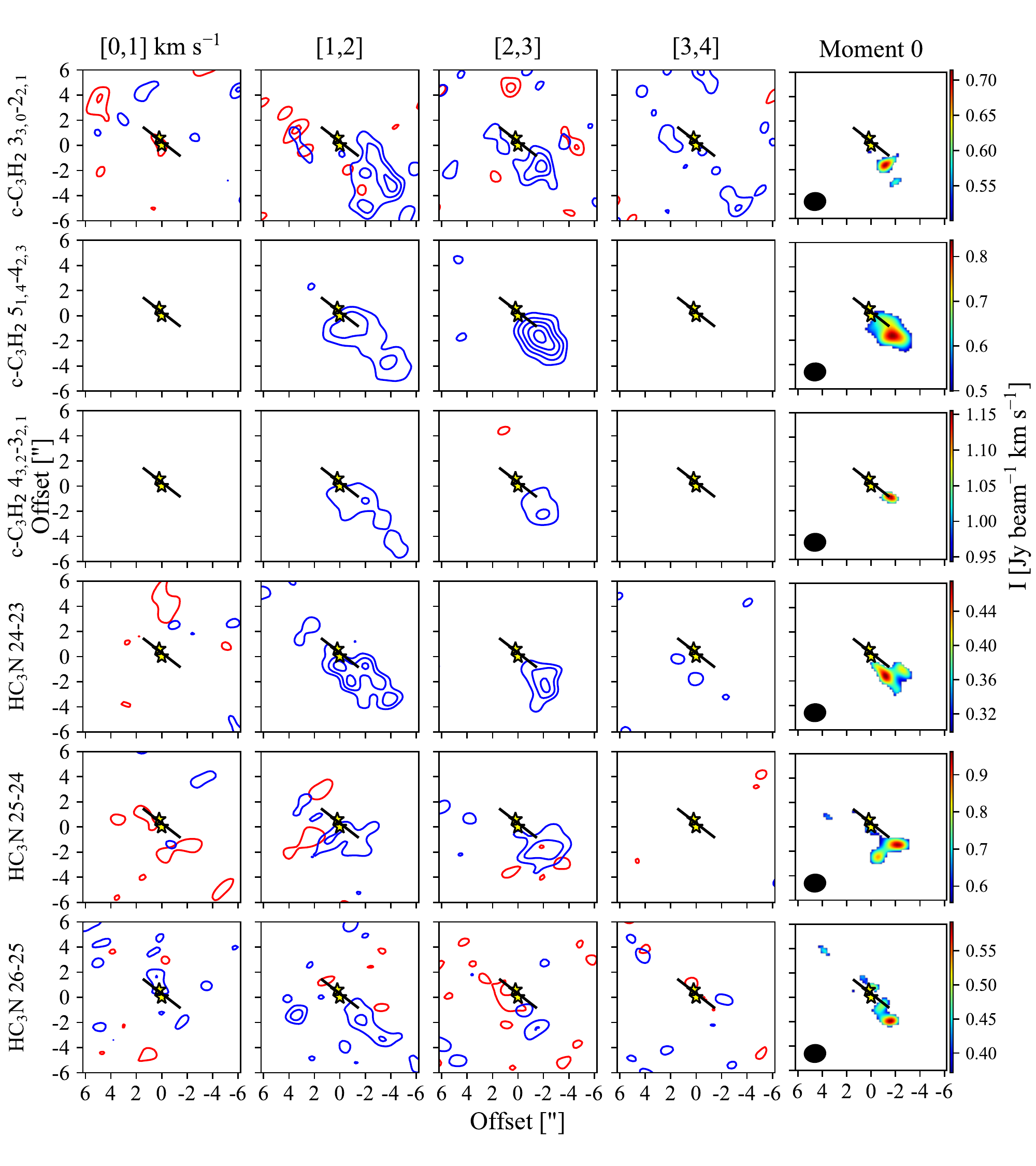}
      \caption[]{\label{fig:C3H2_HC3N}
      Emission of c-C$_{3}$H$_{2}$ and HC$_{3}$N. Channel maps consist of velocity ranges of 1~km~s$^{-1}$ and moment 0 maps are integrated over a velocity range of 8~km~s$^{-1}$. The contours start at 3$\sigma$ and follow a step of 3$\sigma$ and 2$\sigma$ for c-C$_{3}$H$_{2}$ and HC$_{3}$N, respectively. The yellow stars show the position of the sources and the solid segment represents the extent of the circumbinary disc. The synthesised beam is represented by a black filled ellipse in the right panels. 
   }
\end{figure*}

\begin{figure}[t]
   \centering
      \includegraphics[width=.48\textwidth]{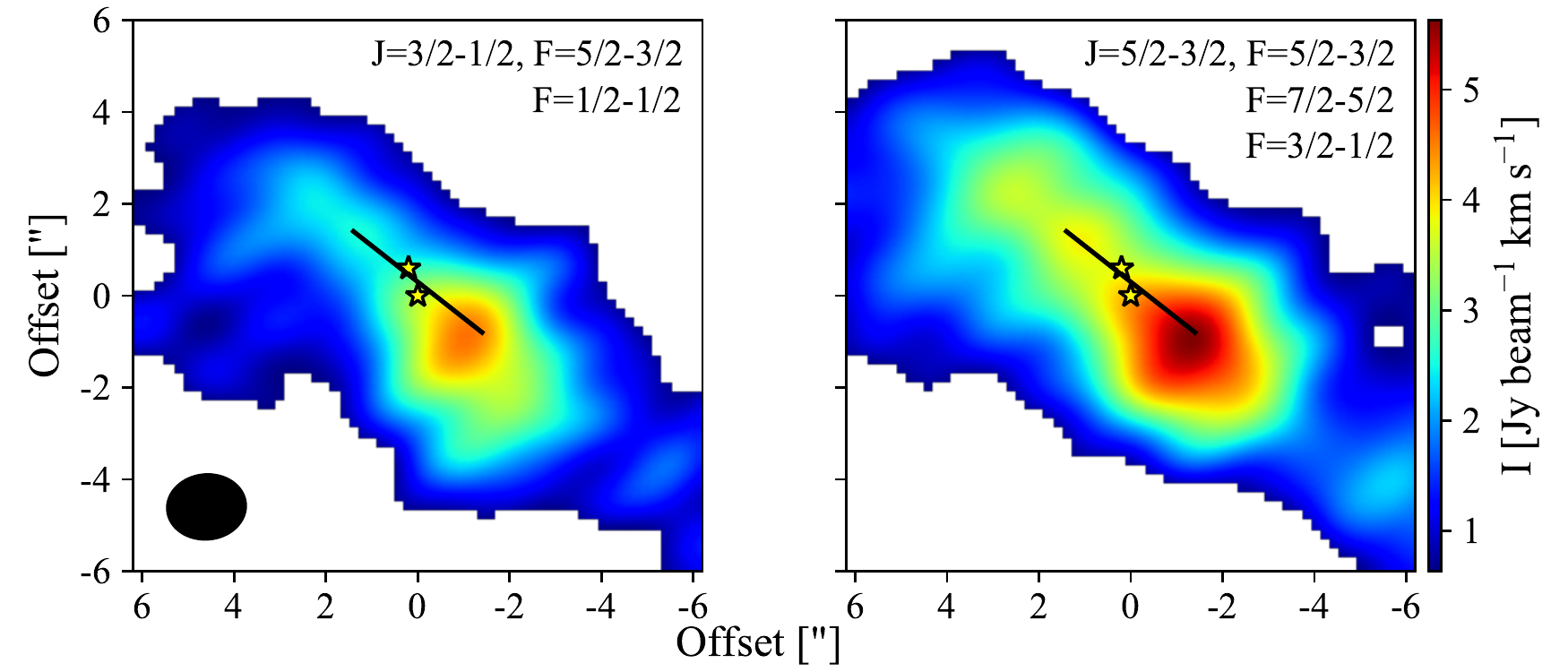}
      \caption[]{\label{fig:CN_combination}
      CN moment 0 maps of the brightest transitions, which are blended with other hyperfine lines. The integration is done over a velocity range of 11 and 8~km~s$^{-1}$ for the left and right panels, respectively. The yellow stars show the position of the sources and the solid segment represents the extent of the circumbinary disc. The synthesised beam is represented by a black filled ellipse in the left panel. 
      }
\end{figure}

Other transitions and less abundant isotopologues of the species discussed in Sect. 3.2 show weaker emission and are presented in Figs.~\ref{fig:SO}, \ref{fig:CS_isotop_H2CO}, and \ref{fig:C3H2_HC3N}. In addition, the brightest CN hyperfine transitions are blended with other CN lines (see Fig.~\ref{fig:spec_CN}). Emission of these transitions is shown in Fig. \ref{fig:CN_combination}.

\section{Non-detections}

Non-detected lines at the 3$\sigma$ level within the covered molecular transitions are listed in Table~\ref{table:non_detected_molecules}, along with their spectroscopic parameters.

\begin{table*}[t]
        \caption{Parameters of the non-detected molecular transitions.}
        \label{table:non_detected_molecules}
        \centering
        \begin{tabular}{l l c c r l}
                \hline\hline
                Species         		& Transition    							& Frequency $^{(a)}$& \textit{A$_{ij}$} $^{(a)}$      	& \textit{E$_\mathrm{u}$} $^{(a)}$	& \textit{n$_\mathrm{crit}$} $^{(b)}$	\\
                                			&                       						& (GHz)                 	& (s$^{-1}$)                      		&  (K)                                          	& (cm$^{-3}$)                             	\\
                \hline
                \multicolumn{6}{c}{CO- species}                                                                                                                                                                                                                							\\
                \hline
                p-H$_{2}$CO           	& 11$_{2,9}$$-$12$_{0,12}$                         	& 215.97616       	& 3.16 $\times$ 10$^{-7}$       	& 280   						& 3.3 $\times$ 10$^{4}$                 	\\
                o-H$_{2}$CO           	& 9$_{1,8}$$-$9$_{1,9}$                                    	& 216.56865       	& 7.24 $\times$ 10$^{-6}$       	& 174   						& 2.1 $\times$ 10$^{5}$         		\\
                CH$_{3}$OH      	& 5$_{0,5}$$-$4$_{0,4}$                               	& 241.70022       	& 6.03 $\times$ 10$^{-5}$         	& 48        						& 1.7 $\times$ 10$^{6}$           		\\
                CH$_{3}$OH           	& 5$_{0,5}$$-$4$_{0,4}$$^{++}$                    	& 241.79143       	& 6.03 $\times$ 10$^{-5}$        	& 35            					& 4.6 $\times$ 10$^{5}$           		\\
                CH$_{3}$OH            	& 5$_{1,4}$$-$4$_{1,3}$                            	& 241.87907       	& 6.03 $\times$ 10$^{-5}$      	& 56            					& 1.5 $\times$ 10$^{7}$           		\\
                CH$_{3}$OH            	& 5$_{1,4}$$-$4$_{1,3}$$^{--}$                        	& 241.91583       	& 6.03 $\times$ 10$^{-5}$        	& 50            					& 5.5 $\times$ 10$^{7}$           		\\
                \hline
                \multicolumn{6}{c}{S- species}                                                                                                                                                                                                                          						\\
                \hline
                SO                        	& 1$_{2}$$-$2$_{1}$                               		& 236.45229       	& 1.41 $\times$ 10$^{-6}$       	& 16            					& 4.0 $\times$ 10$^{4}$ $^{(c)}$    	\\
                SO$_{2}$                 	& 11$_{1,11}$$-$10$_{0,10}$                            	& 221.96522       	& 1.15 $\times$ 10$^{-4}$        	& 60            					& 8.6 $\times$ 10$^{6}$ $^{(d)}$    	\\
                SO$_{2}$                  	& 12$_{3,9}$$-$12$_{2,10}$                    		& 237.06883       	& 1.15 $\times$ 10$^{-4}$      	& 60            					& 1.1 $\times$ 10$^{8}$ $^{(d)}$    	\\
                SO$_{2}$                  	& 5$_{2,4}$$-$4$_{1,3}$                                  	& 241.61580       	& 8.51 $\times$ 10$^{-5}$        	& 24            					& 9.4 $\times$ 10$^{6}$ $^{(d)}$    	\\
                SO$_{2}$                 	& 14$_{0,14}$$-$13$_{1,13}$                        	& 244.25422       	& 1.62 $\times$ 10$^{-4}$     	& 94            					& 1.2 $\times$ 10$^{7}$ $^{(d)}$    	\\
                OCS                        	& \textit{J}=18$-$17                                          	& 218.90336       	& 3.02 $\times$ 10$^{-5}$      	& 100   						& 8.2 $\times$ 10$^{5}$                	\\
                OCS                         	& \textit{J}=19$-$18                                          	& 231.06099       	& 3.55 $\times$ 10$^{-5}$        	& 111   						& 1.3 $\times$ 10$^{6}$                	\\
                OCS                         	& \textit{J}=20$-$19                                           	& 243.21804       	& 4.17 $\times$ 10$^{-5}$        	& 123  						& 3.5 $\times$ 10$^{5}$                	\\      
                p-H$_{2}$S              	& 2$_{2,0}$$-$2$_{1,1}$                                	& 216.71044       	& 4.90 $\times$ 10$^{-5}$      	& 84            					& 1.1 $\times$ 10$^{6}$           		\\
                HCS                         	& 6$_{0,6}$$-$5$_{0,5}$                               	& 241.68929       	& 1.20 $\times$ 10$^{-5}$       	& 41            					&                                               		\\
                HSC                         	& 6$_{1,6}$$-$5$_{1,5}$                                   	& 239.05112       	& 4.27 $\times$ 10$^{-4}$       	& 53            					&                                               		\\
                HSC                         	& 6$_{2,5}$$-$5$_{2,4}$                                  	& 243.32134       	& 4.17 $\times$ 10$^{-4}$        	& 94            					&                                               		\\
                HSC                          	& 6$_{0,6}$$-$5$_{0,5}$                               	& 243.36604       	& 4.68 $\times$ 10$^{-4}$     	& 41            					&                                               		\\
                HSC                         	& 6$_{2,4}$$-$5$_{2,3}$                                  	& 243.52763       	& 4.17 $\times$ 10$^{-4}$       	& 94            					&                                               		\\
                H$_{2}$CS               	& 7$_{1,7}$$-$6$_{1,6}$                           		& 236.72702       	& 1.91 $\times$ 10$^{-4}$      	& 59            					& 5.5 $\times$ 10$^{6}$           		\\
                H$_{2}$CS               	& 7$_{0,7}$$-$6$_{0,6}$                               	& 240.26687       	& 2.04 $\times$ 10$^{-4}$        	& 46            					& 2.0 $\times$ 10$^{7}$           		\\
                H$_{2}$CS               	& 7$_{2,6}$$-$6$_{2,5}$                                 	& 240.38205       	& 1.91 $\times$ 10$^{-4}$      	& 99            					& 4.0 $\times$ 10$^{6}$           		\\
                H$_{2}$CS               	& 7$_{2,5}$$-$6$_{2,4}$                              	& 240.54907       	& 1.91 $\times$ 10$^{-4}$     	& 99            					& 6.4 $\times$ 10$^{6}$           		\\
                H$_{2}$CS               	& 7$_{1,6}$$-$6$_{1,5}$                          		& 244.04850       	& 2.09 $\times$ 10$^{-4}$         	& 60            					& 4.1 $\times$ 10$^{6}$           		\\
                \hline
                \multicolumn{6}{c}{Carbon-chain species}                                                                                                                                                                                                                					\\
                \hline
                N$_{2}$D$^{+}$  	& 3$-$2                                                          	& 231.32186       	& 7.08 $\times$ 10$^{-4}$        	& 22            					& 2.1 $\times$ 10$^{6}$ $^{(e)}$    	\\
                H$_{2}$CN               	& 3$_{0,3}$$-$2$_{0,2}$                              	& 219.85186       	& 3.39 $\times$ 10$^{-4}$         	& 21            					&                                               		\\
                H$_{2}$CN               	& 3$_{1,2}$$-$2$_{1,1}$                                	& 227.43700       	& 3.39 $\times$ 10$^{-4}$         	& 34            					&                                               		\\
                c-C$_{3}$H$_{2}$    	& 8$_{2,6}$$-$8$_{1,7}$ (para)                       	& 218.44883       	& 1.48 $\times$ 10$^{-4}$         	& 87            					& 6.7 $\times$ 10$^{7}$           		\\
                c-C$_{3}$H$_{2}$    	& 8$_{3,6}$$-$8$_{2,7}$ (ortho)                      	& 218.44944       	& 1.48 $\times$ 10$^{-4}$         	& 87            					& 3.2 $\times$ 10$^{8}$          		\\
                c-C$_{3}$H$_{2}$    	& 3$_{2,1}$$-$2$_{1,2}$ (ortho)                       	& 244.22215       	& 5.89 $\times$ 10$^{-5}$         	& 18            					& 1.2 $\times$ 10$^{7}$           		\\
                l-C$_{3}$H$_{2}$     	& 11$_{1,11}$$-$10$_{1,10}$                        	& 226.54857       	& 1.07 $\times$ 10$^{-3}$         	& 79            					&                                                 	\\
                l-C$_{3}$H$_{2}$     	& 11$_{0,11}$$-$10$_{0,10}$                            	& 228.60835       	& 1.12 $\times$ 10$^{-3}$         	& 66            					&                                                 	\\
                l-C$_{3}$H$_{2}$     	& 11$_{1,10}$$-$10$_{1,9}$                            	& 230.77801       	& 1.15 $\times$ 10$^{-3}$        	& 80            					&                                                 	\\
                C$_{4}$H                  	& \textit{N}=23$-$22, \textit{J}=47/2$-$45/2    	& 218.83701       	& 4.47 $\times$ 10$^{-5}$         	& 126   						&                                               		\\
                C$_{4}$H                 	& \textit{N}=23$-$22, \textit{J}=45/2$-$43/2      	& 218.87537       	& 4.47 $\times$ 10$^{-5}$         	& 126   						&                                               		\\
                C$_{4}$H               	& \textit{N}=24$-$23, \textit{J}=49/2$-$47/2      	& 228.34862       	& 5.13 $\times$ 10$^{-5}$         	& 137   						&                                               		\\
                C$_{4}$H              	& \textit{N}=24$-$23, \textit{J}=47/2$-$45/2      	& 228.38696       	& 5.13 $\times$ 10$^{-5}$         	& 137   						&                                               		\\
                C$_{4}$H                  	& \textit{N}=25$-$24, \textit{J}=51/2$-$49/2      	& 237.85974       	& 5.75 $\times$ 10$^{-5}$         	& 148   						&                                               		\\
                C$_{4}$H             	& \textit{N}=25$-$24, \textit{J}=49/2$-$47/2     	& 237.89806       	& 5.75 $\times$ 10$^{-5}$         	& 148   						&                                               		\\
                \hline
                \multicolumn{6}{c}{Others}                                                                                                                                                                                                                                      					\\
                \hline
                SiO                        	& \textit{J}=5$-$4                                               	& 217.10498       	& 5.25 $\times$ 10$^{-4}$       	& 31            					& 2.2 $\times$ 10$^{7}$                	\\
                HDO                         	& 3$_{1,2}$$-$2$_{2,1}$                                  	& 225.89672       	& 1.32 $\times$ 10$^{-5}$      	& 167   						& 7.3 $\times$ 10$^{6}$                	\\
                HDO                          	& 2$_{1,1}$$-$2$_{1,2}$                              	& 241.56155       	& 1.17 $\times$ 10$^{-5}$       	& 95            					& 9.6 $\times$ 10$^{5}$           		\\
                HNCO                   	& 10$_{0,10}$$-$9$_{0,9}$                               	& 219.79827       	& 1.48 $\times$ 10$^{-4}$    	& 58            					& 4.4 $\times$ 10$^{8}$           		\\
                HNCO                    	& 11$_{0,11}$$-$10$_{0,10}$                            	& 241.77403       	& 1.95 $\times$ 10$^{-4}$         	& 70            					& 2.3 $\times$ 10$^{7}$           		\\
                \hline
        \end{tabular}
        \tablefoot{$^{(a)}$  Values from the CDMS database \citep{Muller2001}. $^{(b)}$ Calculated values for a collisional temperature of 30~K and collisional rates from the Leiden Atomic and Molecular Database \citep[LAMDA;][]{Schoier2005}.$^{(c)}$ For a collisional temperature of 60~K.$^{(d)}$ For a collisional temperature of 100~K. $^{(e)}$ Calculated for N$_{2}$H$^{+}$. The collisional rates of specific species listed here and in Table~\ref{table:detected_molecules} were taken from the following sources: CO isotopologues from \cite{Yang2010}, H$_{2}$CO and H$_{2}$CS from \cite{Weisenfeld2013}, CH$_{3}$OH from \cite{Rabli2010}, DCO$^{+}$ and N$_{2}$D$^{+}$ from \cite{Flower1999}, DCN and DNC from \cite{Dumouchel2010}, CS isotopologues and SO from \cite{Lique2006}, SO$_{2}$ from \cite{Balanca2016}, OCS from \cite{Green1978}, H$_{2}$S from \cite{Daniel2011}, CN from \cite{Kalugina2015}, C$_{3}$H$_{2}$ from \cite{Chandra2000}, HC$_{3}$N from \cite{Faure2016}, SiO from \cite{Dayou2006}, HDO from \cite{Faure2012}, and HNCO from \cite{Sahnoun2018}.}
\end{table*}

\section{APEX data}

\begin{figure}[t]
   \centering
      \includegraphics[width=.48\textwidth]{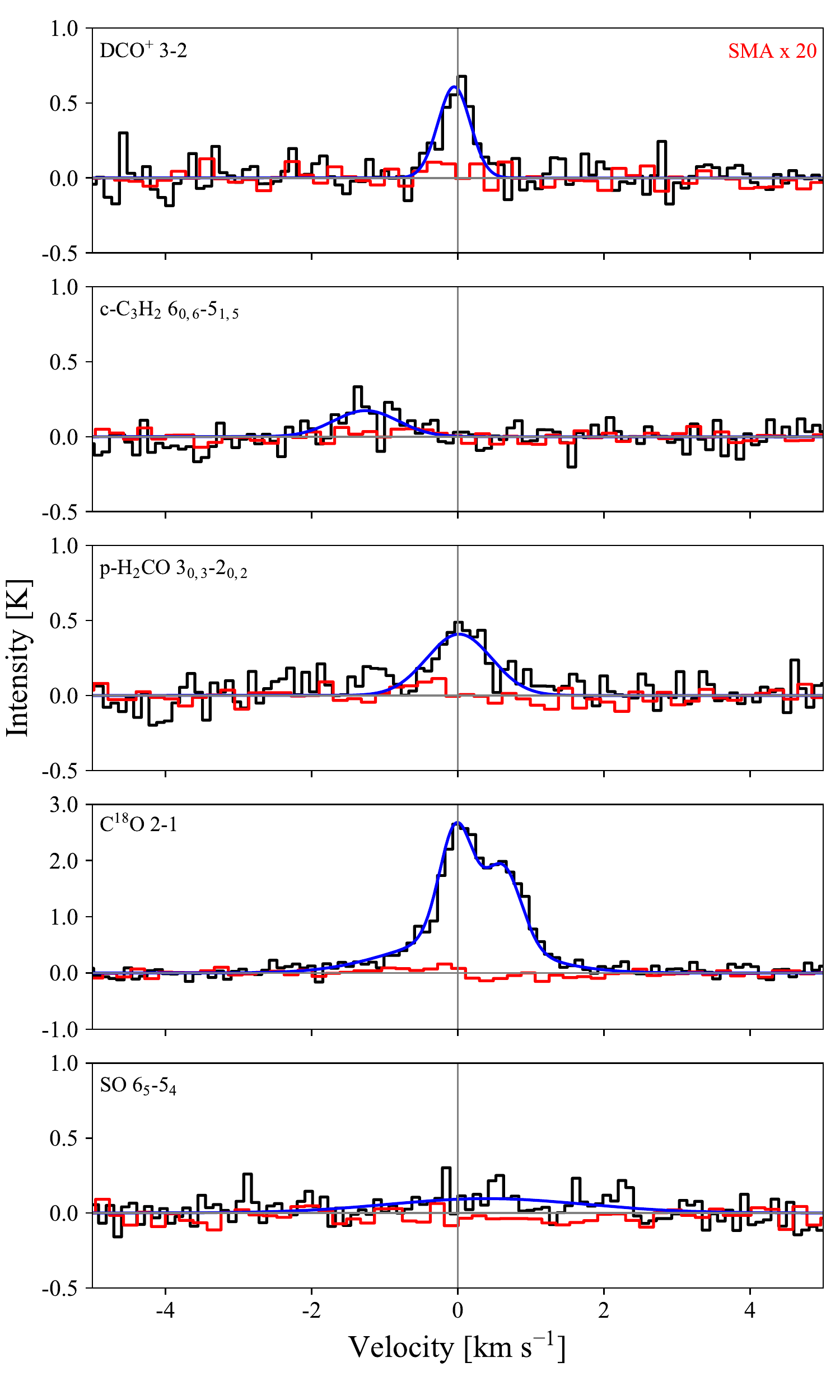}
      \caption[]{\label{fig:APEX_SMA}
      Spectra from APEX data (black) and convolved SMA data with a 29$\arcsec$ beam (red). For clarity, the SMA spectra are scaled with a factor of 20. The blue curve represents a Gaussian fit and, for C$^{18}$O, the curve shows the sum of three Gaussian components. 
      }
\end{figure}

The system IRS 67 was observed with APEX \citep{Lindberg2017} around 220~GHz and with an angular resolution of 29$\arcsec$. In order to compare SMA with APEX data, the SMA observations are convolved with a 29$\arcsec$ beam and the intensity unit (Jy~beam$^{-1}$) is converted into Kelvin. Figure~\ref{fig:APEX_SMA} shows a comparison between the two datasets for five transitions, highlighting the strong filtering-out of extended emission by the interferometer. The APEX and SMA intensities are presented in Table~\ref{table:APEX}, together with the percentage of emission that has been filtered out.

\begin{table}[t]
        \caption{Comparison between APEX \citep{Lindberg2017} and convolved SMA intensities.}
        \label{table:APEX}
        \centering
        \begin{tabular}{l c c c}
                \hline\hline
                Molecular transition                                            	& \multicolumn{2}{c}{Intensity (K~km~s$^{-1}$)}       	& Filtering ($\%$) $^{(a)}$         	\\
                \cline{2-3}
                                                                                        	& APEX                    & SMA                                      	&                                              	\\
                \hline
                DCO$^{+}$ \textit{J}=3$-$2                              	& 0.317           		& 0.0125                               	& \quad96                               	\\
                c-C$_{3}$H$_{2}$ 6$_{0,6}$$-$5$_{1,5}$          & 0.177         		& 0.0137                            	& \quad92                                	\\      
                p-H$_{2}$CO 3$_{0,3}$$-$2$_{0,2}$                 	& 0.431           		& 0.0036                              	& \quad99                                 	\\
                C$^{18}$O \textit{J}=2$-$1                              	& 3.198           		& 0.0093                             	& > 99                                    	\\
                SO 6$_{5}$$-$5$_{4}$                                    	& 0.316           		& 0.0003                               	& > 99                                    	\\
                \hline
        \end{tabular}
        \tablefoot{$^{(a)}$ Percentage of the emission filtered-out by the interferometer.}
\end{table}

\end{appendix}

\bibliographystyle{aa} 
\bibliography{References}

\begin{thebibliography}{70}
\expandafter\ifx\csname natexlab\endcsname\relax\def\natexlab#1{#1}\fi

\bibitem[{{Aikawa} {et~al.}(2018){Aikawa}, {Furuya}, {Hincelin}, \&
  {Herbst}}]{Aikawa2018}
{Aikawa}, Y., {Furuya}, K., {Hincelin}, U., \& {Herbst}, E. 2018, \apj, 855,
  119

\bibitem[{{Aikawa} \& {Herbst}(2001)}]{Aikawa2001}
{Aikawa}, Y. \& {Herbst}, E. 2001, \aap, 371, 1107

\bibitem[{{ALMA Partnership} {et~al.}(2015){ALMA Partnership}, {Brogan},
  {P{\'e}rez}, {Hunter}, {Dent}, {Hales}, {Hills}, {Corder}, {Fomalont},
  {Vlahakis}, {Asaki}, {Barkats}, {Hirota}, {Hodge}, {Impellizzeri}, {Kneissl},
  {Liuzzo}, {Lucas}, {Marcelino}, {Matsushita}, {Nakanishi}, {Phillips},
  {Richards}, {Toledo}, {Aladro}, {Broguiere}, {Cortes}, {Cortes}, {Espada},
  {Galarza}, {Garcia-Appadoo}, {Guzman-Ramirez}, {Humphreys}, {Jung}, {Kameno},
  {Laing}, {Leon}, {Marconi}, {Mignano}, {Nikolic}, {Nyman}, {Radiszcz},
  {Remijan}, {Rod{\'o}n}, {Sawada}, {Takahashi}, {Tilanus}, {Vila Vilaro},
  {Watson}, {Wiklind}, {Akiyama}, {Chapillon}, {de Gregorio-Monsalvo}, {Di
  Francesco}, {Gueth}, {Kawamura}, {Lee}, {Nguyen Luong}, {Mangum}, {Pietu},
  {Sanhueza}, {Saigo}, {Takakuwa}, {Ubach}, {van Kempen}, {Wootten},
  {Castro-Carrizo}, {Francke}, {Gallardo}, {Garcia}, {Gonzalez}, {Hill},
  {Kaminski}, {Kurono}, {Liu}, {Lopez}, {Morales}, {Plarre}, {Schieven},
  {Testi}, {Videla}, {Villard}, {Andreani}, {Hibbard}, \&
  {Tatematsu}}]{ALMA2015}
{ALMA Partnership}, {Brogan}, C.~L., {P{\'e}rez}, L.~M., {et~al.} 2015, \apjl,
  808, L3

\bibitem[{{Artur de la Villarmois} {et~al.}(2019){Artur de la Villarmois},
  {J{\o}rgensen}, {Kristensen}, {Bergin}, {Harsono}, {Sakai}, {van Dishoeck},
  \& {Yamamoto}}]{Artur2019}
{Artur de la Villarmois}, E., {J{\o}rgensen}, J.~K., {Kristensen}, L.~E.,
  {et~al.} 2019, arXiv e-prints, arXiv:1904.13161

\bibitem[{{Artur de la Villarmois} {et~al.}(2018){Artur de la Villarmois},
  {Kristensen}, {J{\o}rgensen}, {Bergin}, {Brinch}, {Frimann}, {Harsono},
  {Sakai}, \& {Yamamoto}}]{Artur2018}
{Artur de la Villarmois}, E., {Kristensen}, L.~E., {J{\o}rgensen}, J.~K.,
  {et~al.} 2018, \aap, 614, A26

\bibitem[{{Bachiller} \& {Perez Gutierrez}(1997)}]{Bachiller1997}
{Bachiller}, R. \& {Perez Gutierrez}, M. 1997, in IAU Symposium, Vol. 182,
  Herbig-Haro Flows and the Birth of Stars, ed. B.~{Reipurth} \& C.~{Bertout},
  153--162

\bibitem[{{Balan{\c c}a} {et~al.}(2016){Balan{\c c}a}, {Spielfiedel}, \&
  {Feautrier}}]{Balanca2016}
{Balan{\c c}a}, C., {Spielfiedel}, A., \& {Feautrier}, N. 2016, \mnras, 460,
  3766

\bibitem[{{Beckwith} \& {Sargent}(1991)}]{Beckwith1991}
{Beckwith}, S.~V.~W. \& {Sargent}, A.~I. 1991, \apj, 381, 250

\bibitem[{{Bergin} {et~al.}(2003){Bergin}, {Calvet}, {D'Alessio}, \&
  {Herczeg}}]{Bergin2003}
{Bergin}, E., {Calvet}, N., {D'Alessio}, P., \& {Herczeg}, G.~J. 2003, \apjl,
  591, L159

\bibitem[{{Bergin} {et~al.}(2016){Bergin}, {Du}, {Cleeves}, {Blake}, {Schwarz},
  {Visser}, \& {Zhang}}]{Bergin2016}
{Bergin}, E.~A., {Du}, F., {Cleeves}, L.~I., {et~al.} 2016, \apj, 831, 101

\bibitem[{{Bontemps} {et~al.}(1996){Bontemps}, {Andre}, {Terebey}, \&
  {Cabrit}}]{Bontemps1996}
{Bontemps}, S., {Andre}, P., {Terebey}, S., \& {Cabrit}, S. 1996, \aap, 311,
  858

\bibitem[{{Ceccarelli}(2004)}]{Ceccarelli2004}
{Ceccarelli}, C. 2004, in Astronomical Society of the Pacific Conference
  Series, Vol. 323, Star Formation in the Interstellar Medium: In Honor of
  David Hollenbach, ed. D.~{Johnstone}, F.~C. {Adams}, D.~N.~C. {Lin}, D.~A.
  {Neufeeld}, \& E.~C. {Ostriker}, 195

\bibitem[{{Chandra} \& {Kegel}(2000)}]{Chandra2000}
{Chandra}, S. \& {Kegel}, W.~H. 2000, \aaps, 142, 113

\bibitem[{{Codella} {et~al.}(2014){Codella}, {Maury}, {Gueth}, {Maret},
  {Belloche}, {Cabrit}, \& {Andr{\'e}}}]{Codella2014}
{Codella}, C., {Maury}, A.~J., {Gueth}, F., {et~al.} 2014, \aap, 563, L3

\bibitem[{{Daniel} {et~al.}(2011){Daniel}, {Dubernet}, \&
  {Grosjean}}]{Daniel2011}
{Daniel}, F., {Dubernet}, M.-L., \& {Grosjean}, A. 2011, \aap, 536, A76

\bibitem[{{Dayou} \& {Balan{\c c}a}(2006)}]{Dayou2006}
{Dayou}, F. \& {Balan{\c c}a}, C. 2006, \aap, 459, 297

\bibitem[{{Dumouchel} {et~al.}(2010){Dumouchel}, {Faure}, \&
  {Lique}}]{Dumouchel2010}
{Dumouchel}, F., {Faure}, A., \& {Lique}, F. 2010, \mnras, 406, 2488

\bibitem[{{Dunham} {et~al.}(2015){Dunham}, {Allen}, {Evans},
  {Broekhoven-Fiene}, {Cieza}, {Di Francesco}, {Gutermuth}, {Harvey},
  {Hatchell}, {Heiderman}, {Huard}, {Johnstone}, {Kirk}, {Matthews}, {Miller},
  {Peterson}, \& {Young}}]{Dunham2015}
{Dunham}, M.~M., {Allen}, L.~E., {Evans}, II, N.~J., {et~al.} 2015, \apjs, 220,
  11

\bibitem[{{Dutrey} {et~al.}(1997){Dutrey}, {Guilloteau}, \&
  {Bachiller}}]{Dutrey1997}
{Dutrey}, A., {Guilloteau}, S., \& {Bachiller}, R. 1997, \aap, 325, 758

\bibitem[{{Faure} {et~al.}(2016){Faure}, {Lique}, \& {Wiesenfeld}}]{Faure2016}
{Faure}, A., {Lique}, F., \& {Wiesenfeld}, L. 2016, \mnras, 460, 2103

\bibitem[{{Faure} {et~al.}(2012){Faure}, {Wiesenfeld}, {Scribano}, \&
  {Ceccarelli}}]{Faure2012}
{Faure}, A., {Wiesenfeld}, L., {Scribano}, Y., \& {Ceccarelli}, C. 2012,
  \mnras, 420, 699

\bibitem[{{Flower}(1999)}]{Flower1999}
{Flower}, D.~R. 1999, \mnras, 305, 651

\bibitem[{{Green} \& {Chapman}(1978)}]{Green1978}
{Green}, S. \& {Chapman}, S. 1978, \apjs, 37, 169

\bibitem[{{Gusdorf} {et~al.}(2008{\natexlab{a}}){Gusdorf}, {Cabrit}, {Flower},
  \& {Pineau Des For{\^e}ts}}]{Gusdorf2008b}
{Gusdorf}, A., {Cabrit}, S., {Flower}, D.~R., \& {Pineau Des For{\^e}ts}, G.
  2008{\natexlab{a}}, \aap, 482, 809

\bibitem[{{Gusdorf} {et~al.}(2008{\natexlab{b}}){Gusdorf}, {Pineau Des
  For{\^e}ts}, {Cabrit}, \& {Flower}}]{Gusdorf2008a}
{Gusdorf}, A., {Pineau Des For{\^e}ts}, G., {Cabrit}, S., \& {Flower}, D.~R.
  2008{\natexlab{b}}, \aap, 490, 695

\bibitem[{{Guzm{\'a}n} {et~al.}(2015){Guzm{\'a}n}, {Pety}, {Goicoechea},
  {Gerin}, {Roueff}, {Gratier}, \& {{\"O}berg}}]{Guzman2015}
{Guzm{\'a}n}, V.~V., {Pety}, J., {Goicoechea}, J.~R., {et~al.} 2015, \apjl,
  800, L33

\bibitem[{{Harsono} {et~al.}(2018){Harsono}, {Bjerkeli}, {van der Wiel},
  {Ramsey}, {Maud}, {Kristensen}, \& {J{\o}rgensen}}]{Harsono2018}
{Harsono}, D., {Bjerkeli}, P., {van der Wiel}, M.~H.~D., {et~al.} 2018, Nature
  Astronomy, 2, 646

\bibitem[{{Harsono} {et~al.}(2014){Harsono}, {J{\o}rgensen}, {van Dishoeck},
  {Hogerheijde}, {Bruderer}, {Persson}, \& {Mottram}}]{Harsono2014}
{Harsono}, D., {J{\o}rgensen}, J.~K., {van Dishoeck}, E.~F., {et~al.} 2014,
  \aap, 562, A77

\bibitem[{{Herbst} \& {van Dishoeck}(2009)}]{Herbst2009}
{Herbst}, E. \& {van Dishoeck}, E.~F. 2009, \araa, 47, 427

\bibitem[{{J{\o}rgensen}(2004)}]{Jorgensen2004d}
{J{\o}rgensen}, J.~K. 2004, \aap, 424, 589

\bibitem[{{J{\o}rgensen} {et~al.}(2007){J{\o}rgensen}, {Bourke}, {Myers}, {Di
  Francesco}, {van Dishoeck}, {Lee}, {Ohashi}, {Sch{\"o}ier}, {Takakuwa},
  {Wilner}, \& {Zhang}}]{Jorgensen2007}
{J{\o}rgensen}, J.~K., {Bourke}, T.~L., {Myers}, P.~C., {et~al.} 2007, \apj,
  659, 479

\bibitem[{{J{\o}rgensen} {et~al.}(2011){J{\o}rgensen}, {Bourke}, {Nguyen
  Luong}, \& {Takakuwa}}]{Jorgensen2011}
{J{\o}rgensen}, J.~K., {Bourke}, T.~L., {Nguyen Luong}, Q., \& {Takakuwa}, S.
  2011, \aap, 534, A100

\bibitem[{{J{\o}rgensen} {et~al.}(2004){J{\o}rgensen}, {Sch{\"o}ier}, \& {van
  Dishoeck}}]{Jorgensen2004c}
{J{\o}rgensen}, J.~K., {Sch{\"o}ier}, F.~L., \& {van Dishoeck}, E.~F. 2004,
  \aap, 416, 603

\bibitem[{{J{\o}rgensen} {et~al.}(2005){J{\o}rgensen}, {Sch{\"o}ier}, \& {van
  Dishoeck}}]{Jorgensen2005b}
{J{\o}rgensen}, J.~K., {Sch{\"o}ier}, F.~L., \& {van Dishoeck}, E.~F. 2005,
  \aap, 437, 501

\bibitem[{{J{\o}rgensen} {et~al.}(2016){J{\o}rgensen}, {van der Wiel},
  {Coutens}, {Lykke}, {M{\"u}ller}, {van Dishoeck}, {Calcutt}, {Bjerkeli},
  {Bourke}, {Drozdovskaya}, {Favre}, {Fayolle}, {Garrod}, {Jacobsen},
  {{\"O}berg}, {Persson}, \& {Wampfler}}]{Jorgensen2016}
{J{\o}rgensen}, J.~K., {van der Wiel}, M.~H.~D., {Coutens}, A., {et~al.} 2016,
  \aap, 595, A117

\bibitem[{{Kalugina} \& {Lique}(2015)}]{Kalugina2015}
{Kalugina}, Y. \& {Lique}, F. 2015, \mnras, 446, L21

\bibitem[{{Kastner} {et~al.}(2018){Kastner}, {Qi}, {Dickson-Vandervelde},
  {Hily-Blant}, {Forveille}, {Andrews}, {Gorti}, {{\"O}berg}, \&
  {Wilner}}]{Kastner2018}
{Kastner}, J.~H., {Qi}, C., {Dickson-Vandervelde}, D.~A., {et~al.} 2018, \apj,
  863, 106

\bibitem[{{Lefloch} {et~al.}(2018){Lefloch}, {Bachiller}, {Ceccarelli},
  {Cernicharo}, {Codella}, {Fuente}, {Kahane}, {L{\'o}pez-Sepulcre}, {Tafalla},
  {Vastel}, {Caux}, {Gonz{\'a}lez-Garc{\'{\i}}a}, {Bianchi}, {G{\'o}mez-Ruiz},
  {Holdship}, {Mendoza}, {Ospina-Zamudio}, {Podio}, {Qu{\'e}nard}, {Roueff},
  {Sakai}, {Viti}, {Yamamoto}, {Yoshida}, {Favre}, {Monfredini},
  {Quiti{\'a}n-Lara}, {Marcelino}, {Boechat-Roberty}, \&
  {Cabrit}}]{Lefloch2018}
{Lefloch}, B., {Bachiller}, R., {Ceccarelli}, C., {et~al.} 2018, \mnras, 477,
  4792

\bibitem[{{Lindberg} {et~al.}(2017){Lindberg}, {Charnley}, {J{\o}rgensen},
  {Cordiner}, \& {Bjerkeli}}]{Lindberg2017}
{Lindberg}, J.~E., {Charnley}, S.~B., {J{\o}rgensen}, J.~K., {Cordiner}, M.~A.,
  \& {Bjerkeli}, P. 2017, \apj, 835, 3

\bibitem[{{Lique} {et~al.}(2006){Lique}, {Spielfiedel}, \&
  {Cernicharo}}]{Lique2006}
{Lique}, F., {Spielfiedel}, A., \& {Cernicharo}, J. 2006, \aap, 451, 1125

\bibitem[{{Mangum} \& {Wootten}(1993)}]{Mangum1993}
{Mangum}, J.~G. \& {Wootten}, A. 1993, \apjs, 89, 123

\bibitem[{{McClure} {et~al.}(2010){McClure}, {Furlan}, {Manoj}, {Luhman},
  {Watson}, {Forrest}, {Espaillat}, {Calvet}, {D'Alessio}, {Sargent}, {Tobin},
  \& {Chiang}}]{McClure2010}
{McClure}, M.~K., {Furlan}, E., {Manoj}, P., {et~al.} 2010, \apjs, 188, 75

\bibitem[{{McMullin} {et~al.}(2007){McMullin}, {Waters}, {Schiebel}, {Young},
  \& {Golap}}]{McMullin2007}
{McMullin}, J.~P., {Waters}, B., {Schiebel}, D., {Young}, W., \& {Golap}, K.
  2007, in Astronomical Society of the Pacific Conference Series, Vol. 376,
  Astronomical Data Analysis Software and Systems XVI, ed. R.~A. {Shaw},
  F.~{Hill}, \& D.~J. {Bell}, 127

\bibitem[{{Miura} {et~al.}(2017){Miura}, {Yamamoto}, {Nomura}, {Nakamoto},
  {Tanaka}, {Tanaka}, \& {Nagasawa}}]{Miura2017}
{Miura}, H., {Yamamoto}, T., {Nomura}, H., {et~al.} 2017, \apj, 839, 47

\bibitem[{{M{\"u}ller} {et~al.}(2001){M{\"u}ller}, {Thorwirth}, {Roth}, \&
  {Winnewisser}}]{Muller2001}
{M{\"u}ller}, H.~S.~P., {Thorwirth}, S., {Roth}, D.~A., \& {Winnewisser}, G.
  2001, \aap, 370, L49

\bibitem[{{Murillo} {et~al.}(2015){Murillo}, {Bruderer}, {van Dishoeck},
  {Walsh}, {Harsono}, {Lai}, \& {Fuchs}}]{Murillo2015}
{Murillo}, N.~M., {Bruderer}, S., {van Dishoeck}, E.~F., {et~al.} 2015, \aap,
  579, A114

\bibitem[{{Murillo} {et~al.}(2018){Murillo}, {van Dishoeck}, {van der Wiel},
  {J{\o}rgensen}, {Drozdovskaya}, {Calcutt}, \& {Harsono}}]{Murillo2018}
{Murillo}, N.~M., {van Dishoeck}, E.~F., {van der Wiel}, M.~H.~D., {et~al.}
  2018, \aap, 617, A120

\bibitem[{{Natta} {et~al.}(2007){Natta}, {Testi}, {Calvet}, {Henning},
  {Waters}, \& {Wilner}}]{Natta2007}
{Natta}, A., {Testi}, L., {Calvet}, N., {et~al.} 2007, Protostars and Planets
  V, 767

\bibitem[{{{\"O}berg} {et~al.}(2015){{\"O}berg}, {Furuya}, {Loomis}, {Aikawa},
  {Andrews}, {Qi}, {van Dishoeck}, \& {Wilner}}]{Oberg2015}
{{\"O}berg}, K.~I., {Furuya}, K., {Loomis}, R., {et~al.} 2015, \apj, 810, 112

\bibitem[{{{\"O}berg} {et~al.}(2010){{\"O}berg}, {Qi}, {Fogel}, {Bergin},
  {Andrews}, {Espaillat}, {van Kempen}, {Wilner}, \& {Pascucci}}]{Oberg2010}
{{\"O}berg}, K.~I., {Qi}, C., {Fogel}, J.~K.~J., {et~al.} 2010, \apj, 720, 480

\bibitem[{{{\"O}berg} {et~al.}(2011){{\"O}berg}, {Qi}, {Fogel}, {Bergin},
  {Andrews}, {Espaillat}, {Wilner}, {Pascucci}, \& {Kastner}}]{Oberg2011b}
{{\"O}berg}, K.~I., {Qi}, C., {Fogel}, J.~K.~J., {et~al.} 2011, \apj, 734, 98

\bibitem[{{Ortiz-Le{\'o}n} {et~al.}(2017){Ortiz-Le{\'o}n}, {Loinard},
  {Kounkel}, {Dzib}, {Mioduszewski}, {Rodr{\'{\i}}guez}, {Torres},
  {Gonz{\'a}lez-L{\'o}pezlira}, {Pech}, {Rivera}, {Hartmann}, {Boden}, {Evans},
  {Brice{\~n}o}, {Tobin}, {Galli}, \& {Gudehus}}]{OrtizLeon2017}
{Ortiz-Le{\'o}n}, G.~N., {Loinard}, L., {Kounkel}, M.~A., {et~al.} 2017, \apj,
  834, 141

\bibitem[{{Qi}(2001)}]{Qi2001}
{Qi}, C. 2001, PhD thesis, CALIFORNIA INSTITUTE OF TECHNOLOGY

\bibitem[{{Rabli} \& {Flower}(2010)}]{Rabli2010}
{Rabli}, D. \& {Flower}, D.~R. 2010, \mnras, 406, 95

\bibitem[{{Sahnoun} {et~al.}(2018){Sahnoun}, {Wiesenfeld}, {Hammami}, \&
  {Jaidane}}]{Sahnoun2018}
{Sahnoun}, E., {Wiesenfeld}, L., {Hammami}, K., \& {Jaidane}, N. 2018, The
  Journal of Physical Chemistry A, 122, 3004

\bibitem[{{Sakai} \& {Yamamoto}(2013)}]{Sakai2013}
{Sakai}, N. \& {Yamamoto}, S. 2013, Chemical Reviews, 113, 8981

\bibitem[{{Sch{\"o}ier} {et~al.}(2005){Sch{\"o}ier}, {van der Tak}, {van
  Dishoeck}, \& {Black}}]{Schoier2005}
{Sch{\"o}ier}, F.~L., {van der Tak}, F.~F.~S., {van Dishoeck}, E.~F., \&
  {Black}, J.~H. 2005, \aap, 432, 369

\bibitem[{{Sipil{\"a}} {et~al.}(2016){Sipil{\"a}}, {Spezzano}, \&
  {Caselli}}]{Sipila2016}
{Sipil{\"a}}, O., {Spezzano}, S., \& {Caselli}, P. 2016, \aap, 591, L1

\bibitem[{{Testi} {et~al.}(2014){Testi}, {Birnstiel}, {Ricci}, {Andrews},
  {Blum}, {Carpenter}, {Dominik}, {Isella}, {Natta}, {Williams}, \&
  {Wilner}}]{Testi2014}
{Testi}, L., {Birnstiel}, T., {Ricci}, L., {et~al.} 2014, Protostars and
  Planets VI, 339

\bibitem[{{Thi} {et~al.}(2004){Thi}, {van Zadelhoff}, \& {van
  Dishoeck}}]{Thi2004}
{Thi}, W.-F., {van Zadelhoff}, G.-J., \& {van Dishoeck}, E.~F. 2004, \aap, 425,
  955

\bibitem[{{van der Tak} {et~al.}(2000){van der Tak}, {van Dishoeck}, {Evans},
  \& {Blake}}]{vanderTak2000}
{van der Tak}, F.~F.~S., {van Dishoeck}, E.~F., {Evans}, II, N.~J., \& {Blake},
  G.~A. 2000, \apj, 537, 283

\bibitem[{{van Dishoeck}(2006)}]{vanDishoeck2006}
{van Dishoeck}, E.~F. 2006, Proceedings of the National Academy of Science,
  103, 12249

\bibitem[{{van Dishoeck}(2018)}]{vanDishoeck2018}
{van Dishoeck}, E.~F. 2018, in IAU Symposium, Vol. 332, IAU Symposium, ed.
  M.~{Cunningham}, T.~{Millar}, \& Y.~{Aikawa}, 3--22

\bibitem[{{van Zadelhoff} {et~al.}(2003){van Zadelhoff}, {Aikawa},
  {Hogerheijde}, \& {van Dishoeck}}]{vanZadelhoff2003}
{van Zadelhoff}, G.-J., {Aikawa}, Y., {Hogerheijde}, M.~R., \& {van Dishoeck},
  E.~F. 2003, \aap, 397, 789

\bibitem[{{Wiesenfeld} \& {Faure}(2013)}]{Weisenfeld2013}
{Wiesenfeld}, L. \& {Faure}, A. 2013, \mnras, 432, 2573

\bibitem[{{Willacy} \& {Langer}(2000)}]{Willacy2000}
{Willacy}, K. \& {Langer}, W.~D. 2000, \apj, 544, 903

\bibitem[{{Wilson}(1999)}]{Wilson1999}
{Wilson}, T.~L. 1999, Reports on Progress in Physics, 62, 143

\bibitem[{{Yang} {et~al.}(2010){Yang}, {Stancil}, {Balakrishnan}, \&
  {Forrey}}]{Yang2010}
{Yang}, B., {Stancil}, P.~C., {Balakrishnan}, N., \& {Forrey}, R.~C. 2010,
  \apj, 718, 1062

\bibitem[{{Yen} {et~al.}(2015){Yen}, {Koch}, {Takakuwa}, {Ho}, {Ohashi}, \&
  {Tang}}]{Yen2015}
{Yen}, H.-W., {Koch}, P.~M., {Takakuwa}, S., {et~al.} 2015, \apj, 799, 193

\bibitem[{{Yoshida} {et~al.}(2019){Yoshida}, {Sakai}, {Nishimura}, {Tokudome},
  {Watanabe}, {Sakai}, {Takano}, \& {Yamamoto}}]{Yoshida2019}
{Yoshida}, K., {Sakai}, N., {Nishimura}, Y., {et~al.} 2019, \pasj
  [\eprint[arXiv]{1901.06546}]

\end{thebibliography}

\end{document}